\documentclass[useAMS, usenatbib]{mn2e}
\usepackage{amsmath}
\usepackage{amssymb}
\usepackage{graphicx}  

\title[Formation histories of simulated disc galaxies]{The diverse formation histories of simulated disc galaxies}
\author[M. Aumer et al.]
{Michael Aumer$^{1,2}$ \thanks{E-mail:maumer@mpa-garching.mpg.de (MA)}, Simon D.M. White$^{1}$ and Thorsten Naab$^{1}$\\
$^{1}$Max-Planck-Institut f\"ur Astrophysik, Karl-Schwarzschild-Str. 1, 85748 Garching, Germany\\
$^{2}$Excellence Cluster Universe, Boltzmannstr. 2, 85748 Garching, Germany}

\begin{document}

\date{Accepted 2014 April 24. Received 2014 April 24; in original form 2014 February 20}

\pagerange{\pageref{firstpage}--\pageref{lastpage}} \pubyear{2014}

\maketitle

\label{firstpage}

\begin{abstract}

We analyze the formation histories of 19 galaxies from cosmological smoothed particle hydrodynamics zoom-in resimulations.
We construct mock three-colour images and show that the models reproduce observed trends in the evolution of galaxy
colours and morphologies. However, only a small fraction of galaxies contains bars. Many galaxies go through phases of
central mass growth by in-situ star formation driven by gas-rich mergers or misaligned gas infall. These events lead
to accretion of low-angular momentum gas to the centres and leave imprints on the distributions of $z=0$ stellar circularities,
radii and metallicities as functions of age. Observations of the evolution of structural properties of samples
of disc galaxies at $z=2.5 - 0.0$ infer continuous mass assembly at all radii. Our simulations can only explain this 
if there is a significant contribution from mergers or misaligned infall, as expected in a $\Lambda$CDM universe. Quiescent
merger histories lead to high kinematic disc fractions and inside-out growth, but show little central growth after the last
`destructive' merger at $z>1.5$. For sufficiently strong feedback, as assumed in our models, a moderate amount of merging
does not seem to be a problem for the $z=0$ disc galaxy population, but may rather be a requirement. The average profiles
of simulated disc galaxies agree with observations at $z\ge 1.5$. At $z\le1$, there is too much growth in size
and too little growth in central mass, possibly due to the under-abundance of bars. The discrepancies may partly be caused
by differences between the star formation histories of the simulations and those assumed for observations. 

\end{abstract}

\begin{keywords}
methods: numerical - galaxies:formation - galaxies:evolution  - galaxies:structure;
\end{keywords}

\section{Introduction}

Galaxies with masses similar to that of the Milky Way (MW) are thought to have
been most efficient of all galaxies in turning available baryons into stars \citep{guo}.
At $z=0$, a majority of them are dominated by discs \citep{delgado}. 
It is therefore a key question in astrophysics to understand the formation histories
of MW-like disc galaxies.

Recently, \citet{vD} (D13 hereafter) and \citet{patel} 
(P13 hereafter) presented observational data on the growth and structural evolution
of such galaxies from $z=2.5$ to $z=0$. To achieve this, D13 selected samples
of galaxies in six disjoint redshift intervals at the same cumulative co-moving number density,
assuming that this would reproduce the typical formation history of a $z=0$ galaxy with a stellar mass of $\sim5\times 10^{10}\;M_{\odot}$.
P13 selected samples of galaxies at six redshifts from $z=1.3$ to $z=0$ according to 
their star formation rates (SFRs) and masses under the assumption that a typical selected galaxy stays on the evolving 
main sequence of star formation (e.g. \citealp{karim}) and has a $z=0$ mass of $3.2\times10^{10}\;M_{\odot}$.

\begin{figure*}
\centering
\vspace{-0.5cm}
\includegraphics[width=17.4cm]{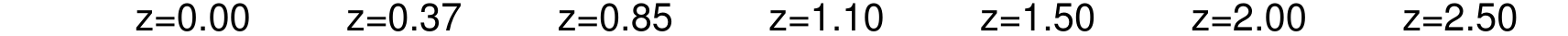}\\
\vspace{-0.15cm}
\hspace{-0.15cm}
\includegraphics[width=1.35cm]{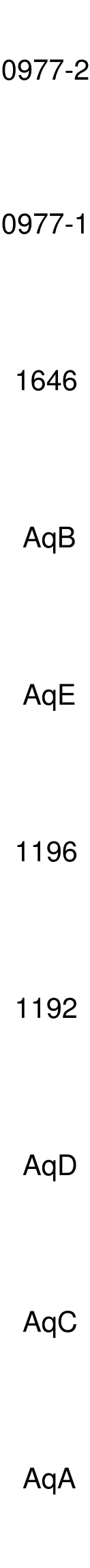}
\hspace{-0.15cm}
\includegraphics[width=16.4cm]{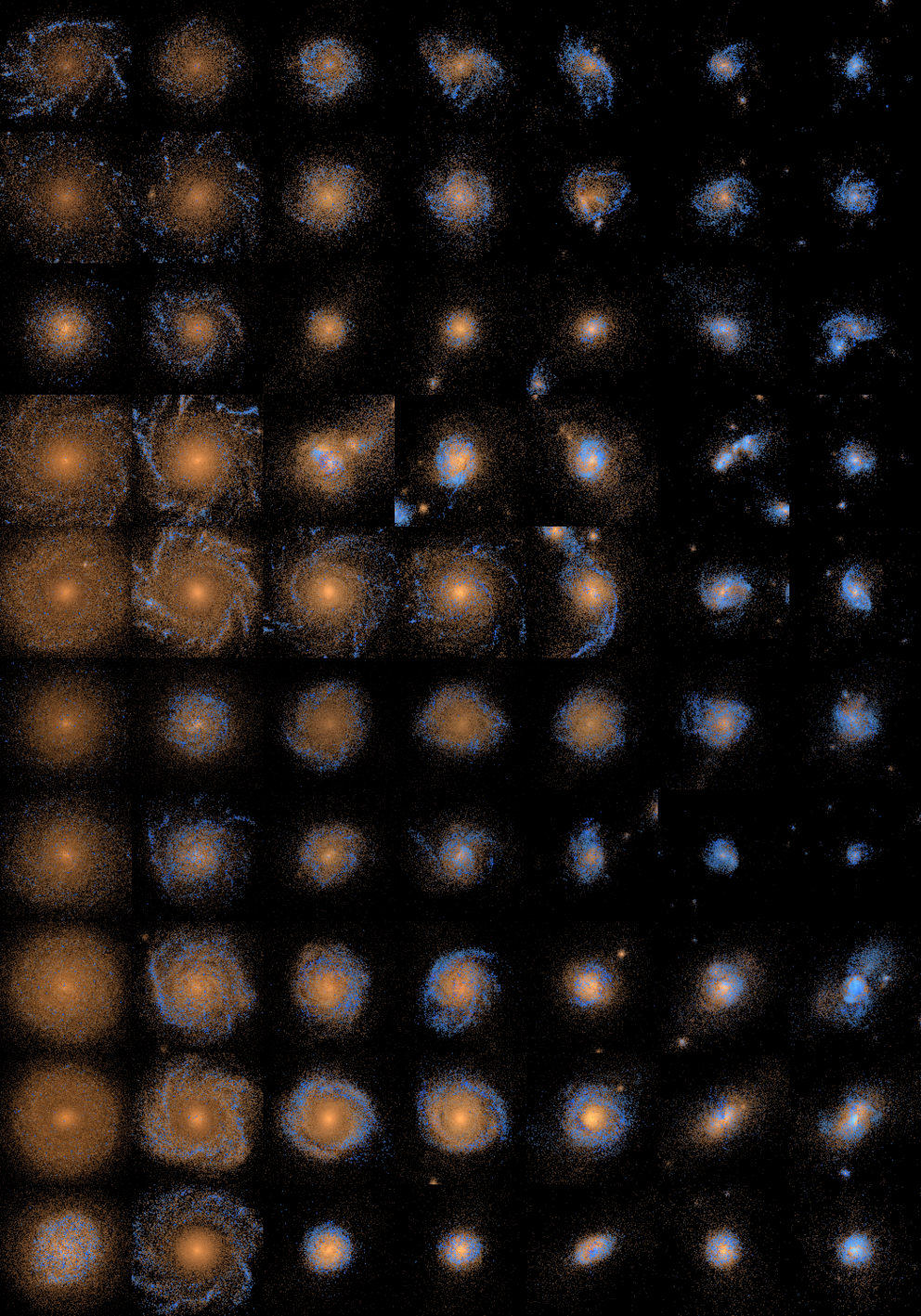}
\caption[Evolution of face-on mock three-colour images 1]
{Evolution of face-on mock three-colour $ugr$-band images of the model galaxies. From right to left redshifts $z=2.5,2.0,1.5,1.1,0.85,0.37,0.0$.
  From top to bottom models 0997-1, 0997-2, 1646, AqB, AqE, 1196, 1192, AqD, AqC, AqA. Each image comprises 40 x 40 kpc.}
\label{evolution1}
\end{figure*}

Both papers concluded that the galaxy populations under consideration grew only mildly in effective radius over the studied time.
Stellar mass growth in the centre occurs at all times with outer growth being only mildly more
efficient at late times. This behaviour differs strongly from the stellar mass growth as observed in massive elliptical galaxies, 
some of which apparently formed dense cores at high redshifts ($z>2$), and which as a population grew strongly in outer mass and effective radius
at later times \citep{vD2, patel2}. This evolutionary path has been explained as a result of the continuous accretion of stars
in dry major and minor mergers \citep{naab, bezanson, hopkins10, oser, oser12, hilz}.

Major mergers destroy thin galactic discs \citep{toomre77}, so disc dominated galaxies are thought to form
in haloes with quiescent merger histories from the cooling of gas from increasingly large radii. This leads
to an inside-out formation process (e.g. \citealp{fall, mo}), for which observational evidence in the form of radial 
stellar age gradients has been presented (e.g. \citealp{califa}). Results from abundance matching support
the idea that most stars in disc galaxies formed in-situ \citep{behroozi,moster, yang}.
Additionally, it has been suggested that for $\Lambda$CDM haloes, feedback processes that preferentially
remove low-angular momentum gas are needed to explain the observed structure of galactic discs \citep{donghia, dutton, brook}.

Several processes that modify the simple inside-out formation picture for disc galaxies are, however, known and can lead to substantial 
star formation (SF) in the central galactic regions. These include clump migration in violently unstable discs \citep{noguchi},
bar-induced gas inflows \citep{atha}, gas-rich minor mergers \citep{barnes}, angular momentum loss due to the 
reorientation of the disc rotation axis \citep{aw, okamoto} and the infall of gas with misaligned angular momentum
\citep{cs09}. 

The fact that a large fraction of local disc galaxies show small bulge fractions or even no evidence for a classical bulge at all
(e.g. \citealp{kormendy}) has, on the one hand, been identified as a possible challenge to hierarchical galaxy formation in a 
$\Lambda$CDM universe (e.g. \citealp{freeman}). On the other hand, it has also been shown that discs can (re)form after gas-rich
mergers \citep{barnes02, sprihern, naab6, robertson}. Observational evidence for this process has been presented in the form
of peculiar morphologies, anomalous kinematics and episodic star formation histories (SFHs) of spiral galaxies at $z\sim 0.5$
(e.g. \citealp{hammer,hammer9,puech}). Additionally, \citet{zavala} have shown that a semi-empirical model for the bulge growth
through mergers in a $\Lambda$CDM universe can correctly reproduce the observed $z=0$ distribution of galactic bulge fractions. 
They have, however, ignored other channels of bulge growth.

In this work we use recent cosmological hydrodynamical simulations by \citet{a13}
to better understand the observed structural evolution of disc galaxies.
Our paper is organized as follows:
In Section 2 we describe the sample of simulated galaxies.
In Section 3 we illustrate the diverse formation histories of the models.
In Section 4 we discuss the influence of mergers and misaligned infall.
In Section 5 we compare to recent observations.
Finally, we conclude in Section 6.

\section{The simulated galaxies}

For our analysis of simulated disc galaxies, we rely on the models recently presented in \citet{a13} (A13 hereafter).
These re-simulations follow the formation of galaxies in a $\Lambda$CDM universe, including dark matter and baryonic physics.
The simulations were run with the TREE-Smoothed Particle Hydrodynamics (SPH) code GADGET-3, last described in \citet{gadget2}.
Models for baryonic physics were presented in \citet{cs05, cs06} and A13.

They include a Schmidt-type SF
law above a threshold density of $n_{\rm th}\sim 3 \; {\rm cm^{-3}}$. 
The comoving gravitational softening lengths for baryonic and dark matter particles are
$\epsilon_{\rm baryons}\sim300\;{\rm pc}$ and $\epsilon_{\rm DM}\sim650\;{\rm pc}$. The particle masses for baryons and
dark matter are in the ranges $m_{\rm baryons}=2.87-7.37 \times10^{5}M_{\odot}$ and $m_{\rm DM}=1.50-3.62 \times10^{6}M_{\odot}$.
The effects of the sub-grid physics implementation and the choices of parameters are in detail discussed in A13.
At the given resolution of the models, lower values for the SF threshold density
$n_{\rm th}$ lead to less efficient feedback and less prominent galactic discs, whereas higher values lead to spurious effects
due to a lack of resolution. The coarse resolution also prevents the formation of sufficiently cold, thin gas discs,
as gas velocity dispersions are limited to values above $\sigma\sim 15\;{\rm kms^{-1}}$ (see also \citealp{house}).

Simulations also include models for  metal production by supernovae (SNe) of types II and Ia and
by asymptotic giant branch stars, and metal-line cooling based on individual element abundances \citep{wiersma}.
Metals can also mix by a scheme modelling their turbulent diffusion. Feedback energy ejected from SNe into the ISM is
split into a kinetic and a thermal part and, additionally, a model for feedback from radiation pressure of massive young stars on the
surrounding ISM is included. The application of an explicit multiphase scheme for gas particles allows the coexistence of
and the exchange of material between a cold, dense phase, from which stars can form, and a diffuse, hot phase.

Our sample includes 19 $z=0$ galaxies, 18 of which were already discussed in A13 and one additional galaxy (AqE), 
which in all studied properties behaves very similarly to the other studied galaxies.
The initial conditions include five Aquarius haloes \citep{aquarius, cs09}, which were specifically selected to be
prime candidates for hosting disc dominated galaxies at $z=0$ and consequently have rather quiescent merger histories.
The remaining initial conditions originate from the sample presented by \citet{oser} and have diverse assembly histories.
Specifically, 15 of the 19 galaxies undergo merger events at $z<1$, which have a destructive effect on the stellar disc
(cf. Figure 13 in A13). At $z=0$, 15 of the 19 galaxies can be regarded as central galaxies, whereas two haloes each host two
galaxies of similar mass, which are about to merge in the near future.

The halo mass range of the model galaxies is $1.7\times 10^{10}<M_{200}/M_{\odot}<2.4\times10^{12}$ and the model galaxies at $z=0$
have stellar masses of $3\times10^{9}<M_{\rm stellar}/M_{\odot}<2\times10^{11}$. A13 showed that the SFHs
and the $z=0$ stellar masses of the model galaxies are in good agreement with recent results from abundance matching
by \citet{moster}.

As was shown in A13, all but one of the model galaxies show significant disc components and are actively star-forming at $z=0$.
They show realistic morphologies in gas and stars, realistic surface brightness profiles and flat circular velocity curves.
The evolution of gas fractions, stellar and gas phase metallicities is also in good agreement with observational constraints.
The model galaxies thus present an interesting sample to be compared to recent observations of structural evolution
of disc galaxies.

\section{Structural evolution of simulated disc galaxies}

We begin our analysis of simulated disc galaxies by creating mock three-colour images of galaxies at various redshifts.
We therefore rotate each galaxy at each considered redshift to face-on orientation and create square pixels with
side lengths of $100-200$ pc. Using the stellar population synthesis models of \citet{bruzual} 
for a \citet{chabrier} initial mass function, we determine $u$, $g$, and 
$r$ band luminosities for all pixels taking into account masses, metallicities and ages of star particles.
We also take into account obscuration effects by calculating gas-phase metal mass distributions along the line-of-sight 
for each pixel from the masses, SPH smoothing lengths and metallicities of gas particles with $T<15000\;{\rm K}$. 
We use these as a proxy for the dust mass distributions. For each star particle, we thus obtain a dust surface density
according to its position along the line-of-sight, which we convert into an extinction factor following the relations
presented in \citet{pei}. We assume that their results for the Milky Way correspond to solar metallicity gas
and replace HI mass, as used in their relation, by cold gas mass. The extinction corrected $u/g/r$ luminosities are then
mapped to blue/green/red colours to create three-colour images.
This procedure was motivated by the images presented by \citet{hopkins}.
Similar image series without extinction corrections were recently also presented by \citet{marinacci} for their
cosmological simulations of disc galaxies.

\begin{figure}
\centering
\includegraphics[width=8.cm]{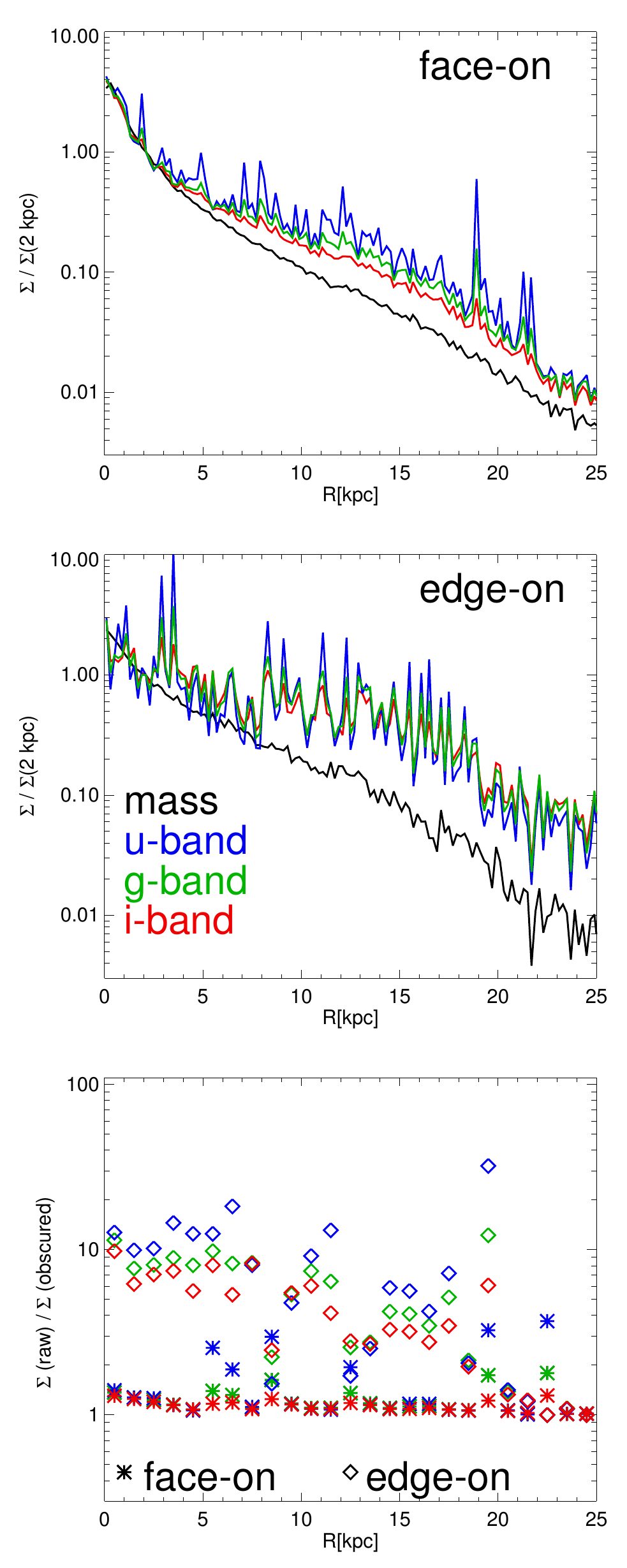}
\caption[Mass and light profiles of model 1192 at $z=0$]
{Mass and light profiles of model 1192 at $z=0$. Top: Face-on profiles for mass (black), $i$ band (red), $g$ band (green) and $u$ band (blue).
 Middle: As above, but for edge-on profiles. The profiles are normalized to intersect at
 $R=2$ kpc. Bottom: The ratio of raw to obscured light as a function of radius $R$ for edge-on (diamonds)
 and face-on (asterisks) profiles in $i$, $g$ and $r$ bands. }
\label{radial-light}
\end{figure}

Images with sizes of 40 x 40 kpc for ten galaxies at redshifts $z=$ 2.5, 2.0, 1.5, 1.1, 0.85, 0.37 and 0.0 are presented in Fig. \ref{evolution1}.
We can compare these images to the observed images presented in Fig. 2 of D13 taking into account
that the observed galaxies show various orientations and are presented in two-colour $u/g$ band images.
As in the observed galaxy sample, the model galaxies change significantly during their evolution and become
redder at redshifts $z<1$ as the specific SFRs decrease with time.
The total sizes of galaxies grow significantly.

The galaxies also undergo morphological changes with spiral structures being common at low $z$, but
less prominent at high $z$. Spiral structure is typically flocculent and there are very few indications of bars.
For several galaxies, the fading of spiral structure due to a decrease in SFR is visible at z=0.
The importance of mergers is indicated by pairs, tidal structures and other irregular morphologies displayed by
several galaxies, preferentially at higher redshifts.

\begin{figure*}
\centering
\includegraphics[width=18cm]{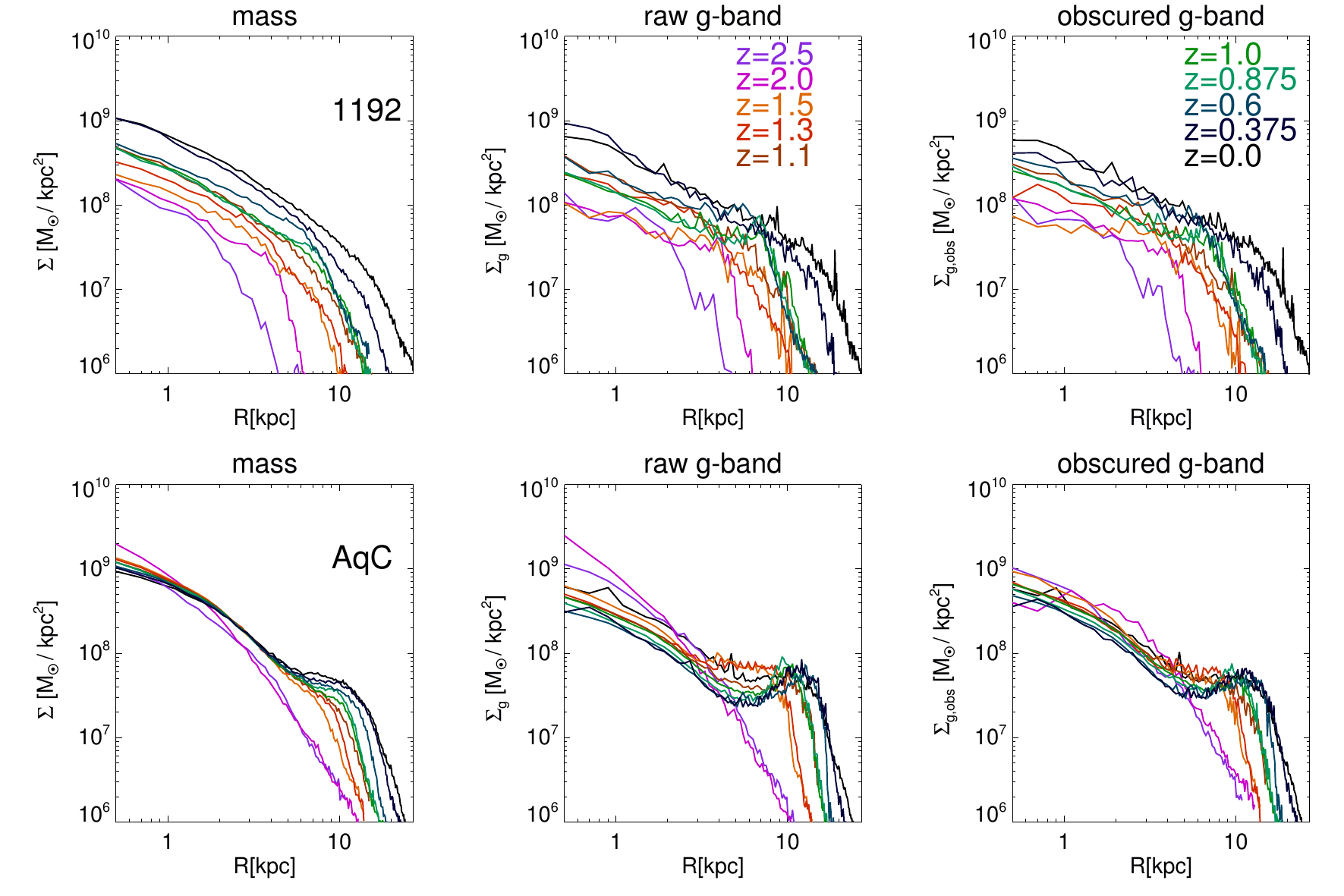}
\caption[Comparison of mass and light profile evolution]
{The evolution of face-on surface mass density (left) and $g$ band surface brightness profiles (raw: middle, obscured: right)
for two models: 1192 (top) and AqC (bottom).}
\label{radial-g}
\end{figure*}

At low $z$, there are several galaxies having red central regions deficient in SF,
and bluer outskirts with higher SFRs, but also a few galaxies with blue, actively star forming centres. 
Some model galaxies display phases of disc growth around a redder centre after a phase of central SF,
but typically discs do not assemble around `naked' bulges. Phases of replenished central SF can also follow
after inside-out growth phases. Moreover, there are galaxies which do not show a strong colour gradient between inner and outer parts.
We conclude, that there is considerable diversity in behaviour and good qualitative agreement between observed and mock images.

One generic problem in comparing structural properties of galaxies in observations and simulations are conversions
from observed surface brightness to stellar mass and from simulated stellar mass to surface brightness. Moreover, D13 and P13 use
different bands for their analysis of structural evolution, bands close to rest-frame $g$ band (D13) and
bands close to rest frame $i$ band (P13) respectively. 
Orientations of galaxies also complicate the picture, as profiles from edge-on and face-on galaxies give different results.
We attempt to address these complications in Fig. \ref{radial-light}.

We create mock images in $u/g/i$ band, as detailed above. This gives us qualitative insight into
the effects involved, but is less sophisticated than fully self-consistent radiative transfer post-processing \citep{jonsson}, 
which would, however, also result in significant parameter uncertainties and large computational costs.

The upper panel of Fig. \ref{radial-light} compares face-on radial profiles of model 1192 at $z=0$ 
for mass (black) and for $u/g/i$ band light (blue/green/red) normalized to their values at $R=2 \; {\rm kpc}$. 
We note that in all bands, the outer disc of the galaxy appears more prominent compared to the central regions
than in the mass profile. The effective scale-length is highest in $b$ band, decreases with increasing wavelength
and is smallest for the mass profile. This is in qualitative agreement with observations that find
 longer disc scale-lengths at shorter wavelengths (e.g. \citealp{macarthur}) and of half-mass radii
that are smaller than optical half-light radii (e.g. \citealp{califa}). The $u$ band, which is a good tracer for
young stars, is also significantly less smooth than the mass profile due to the clumpy nature of SF.
The spikes produced by SF regions are less pronounced at longer wavelengths.
The asterisks in the lower panel of Fig. \ref{radial-light} give the face-on extinction factors predicted by our
procedure. These are typically below 2, only for some SF regions, there are higher values in the $u$ band.

The discrepancies between light and mass profile in 1192 are a bit stronger than average for galaxies at $z=0$.
We find the strongest discrepancies for galaxies with old, red centres surrounded by blue star-forming `rings'.
Hardly any discrepancies are found for the spheroidal model 6782, which has stopped SF after a merger at $z\sim0.4$.
At high $z$, discrepancies are typically less distinct, as radial stellar population gradients are smaller.
Extinction effects are stronger at high $z$ due to higher gas fractions. A full discussion of discrepancies
between light and mass profiles is beyond the scope of this work.

If we redo the exercise for edge-on views and create profiles for a slit of 1 kpc width, the results are
markedly different. The mass profile shows an almost identical exponential decline outside $R\sim 5 \; {\rm kpc}$ as in the face-on profile,
but the central excess is weakened due to projection effects. However, the extinction caused by the significantly larger dust 
column densities severely affects the light profiles, which in all bands are shallower than the mass
profile. The different bands differ primarily in variability for the reasons discussed above.
The extinction factors, as shown by the diamonds in the lower panel of Fig. \ref{radial-light}, are typically of the
order of 3-10 but can be as high as 50 in the $u$ band.

We conclude that both the selection of the colour band and the orientation can have a significant effect on the determination of
structural properties for disc galaxies. We have shown that both $g$ and $i$ band profiles do not accurately trace mass profiles,
with discrepancies being smaller in $i$ band. In our work, we now restrict ourselves to face-on profiles 
and caution that the above findings should be kept in mind. 

In Fig. \ref{radial-g} we depict the structural evolution of two galaxies with very different formation 
histories. We show profiles in mass, raw $g$ band and obscured $g$ band from $z=2.5$ to $z=0$. For the light profiles, we normalize each profile
to the same total `mass' as the corresponding mass profile.

\begin{figure*}
\centering
\includegraphics[width=18cm]{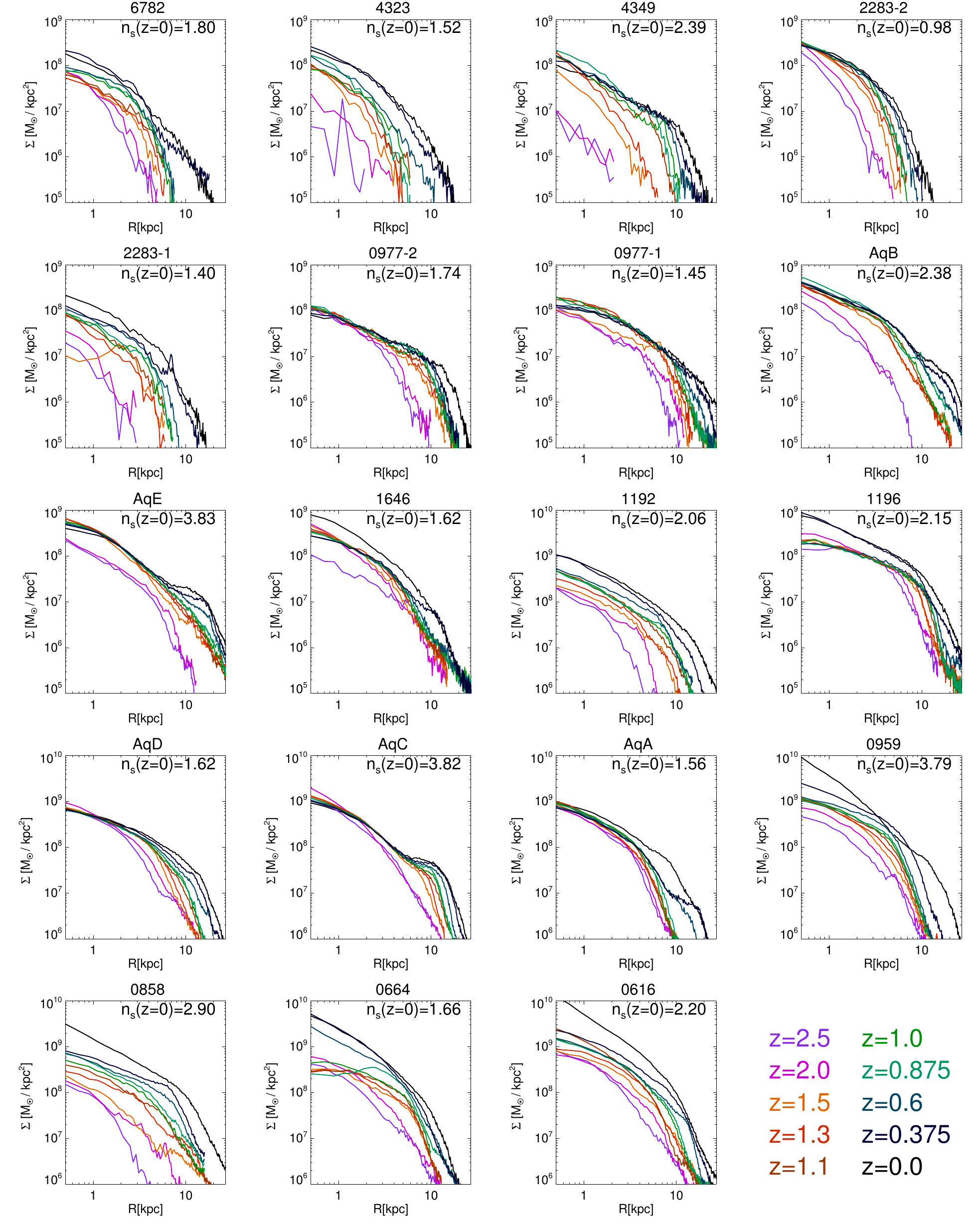}
\caption[Surface mass density evolution for all 19 galaxies]
{The evolution of face-on surface mass density profiles $\Sigma(R)$ from $z=2.5$ to $z=0$ for all 19 model galaxies.
The models are sorted by $z=0$ stellar mass. The most massive galaxy (0616) is on the bottom right. In each panel
we give the Sersic index $n_s$ that we find when fitting the $z=0$ $\Sigma(R)$ profile with a one-component Sersic profile.}
\label{radial-all}
\end{figure*}

The model galaxy 1192 (seventh row in Fig. \ref{evolution1}) experiences its last significant merger at $z\sim 0.5$ and develops
a central bar afterwards. Its stellar mass increases continuously at all radii, it does not experience classical inside-out growth.
When analyzing raw and obscured $g$ band light profiles, we notice that these are more extended (see discussion above)
and less smooth, but the continuous growth at all radii remains clearly visible .

Model galaxy AqC (ninth row in Fig. \ref{evolution1}) 
evolves very differently. Its central mass density only increases at early times and later decreases due to
stellar mass loss. After $z=2$, there is little change in surface density at $R<5 \; {\rm kpc}$. 
Gas infall then happens predominantly in the outer regions, which leads to the formation of a
star-forming ring of varying extent around the dead central region. The galaxy thus displays inside-out formation.
In raw $g$-band light, the 
evolution looks significantly different (see e.g. also the models of \citealp{naabo}). The central surface brightness decreases
continuously as no new stars form and bright stars die early. Taking obscuration into
account, this behaviour is still visible, but it is weaker, as galaxies are richer in gas and dust relative to stars at earlier times.
As it is the site of SF, and thus of bright young stars, the ring is significantly more distinct in surface brightness
than in mass surface density, constituting a local maximum in surface brightness both in the raw and in the obscured light profiles.
If such effects are not accounted for, the inferred structural history would differ significantly
 depending on whether mass or $g$-band light were used.

Having studied the differences between light and mass profiles, we use Fig. \ref{radial-all} to give an overview of the
structural evolution of all 19 model galaxies, restricting ourselves to face-on surface mass density profiles. 
As was discussed in A13, the SFRs at redshifts $z\le 0.5$ of the 
model galaxies with the lowest and highest masses deviate from 
observations. The lower mass models tend to underproduce stars at these 
times, whereas the four highest mass galaxies overproduce stars at low
redshifts (the models are sorted by mass in Fig. \ref{radial-all}). For
the high mass models, this overproduction is visible as the strong mass
growth between $z=0.375$ and $z=0$, which happens predominantly in 
central regions. This central growth disagrees also with recent 
observations by \citet{fang} who find that galaxies with the highest
central mass surface densities usually show low $z=0$ SFRs. One possible 
solution for this problem could be the inclusion of a model for AGN 
feedback to stop the cooling of hot gas onto the galaxies at low $z$
\citep{croton,khala}.

\begin{figure*}
\centering
\includegraphics[width=17.cm]{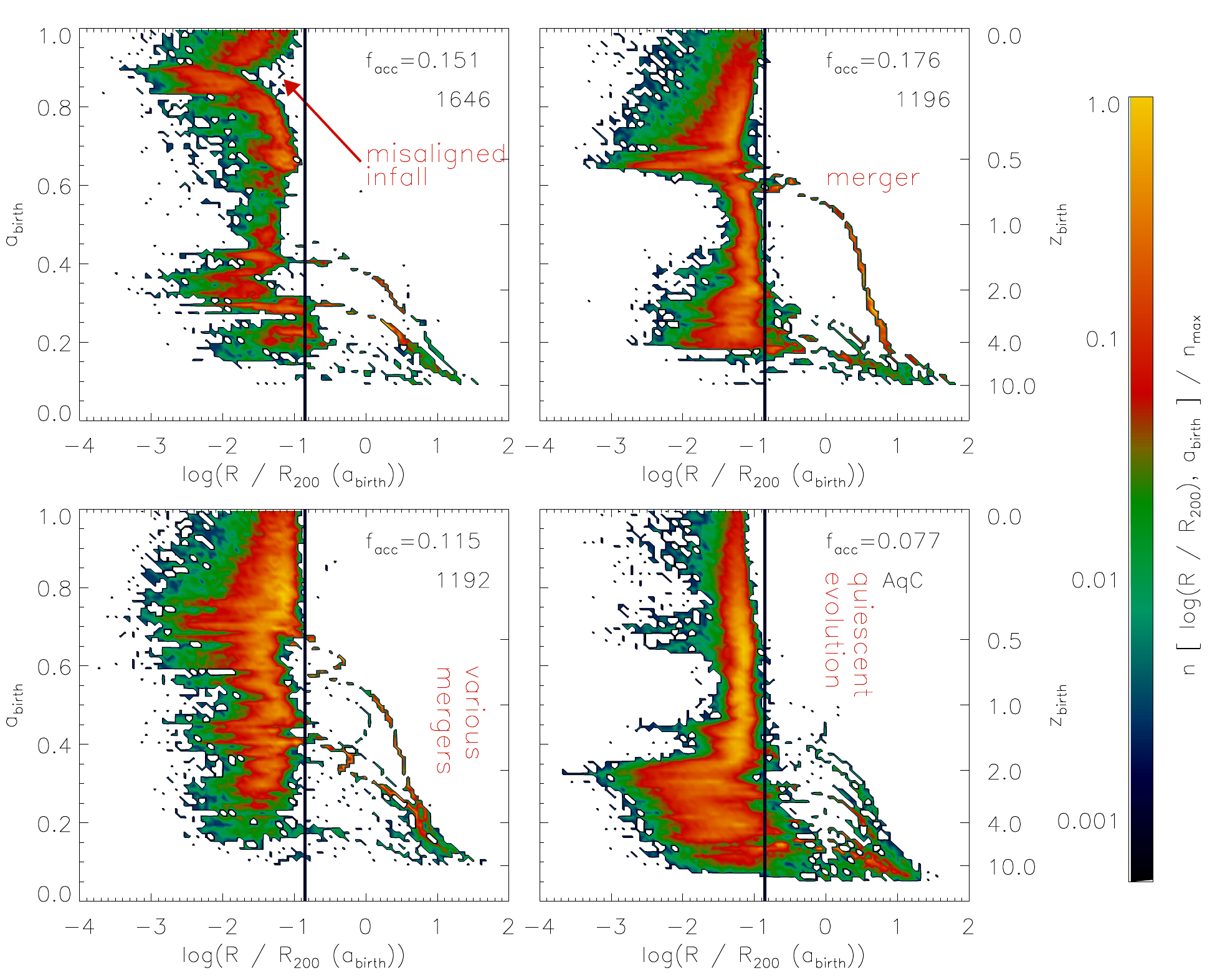}
\caption[Birth radii of galactic stars as a function of birth time]
{The distribution of birth radii of stars which are part of the galaxy at $z=0$ in units of the virial radius $R_{200}(a_{\rm birth})$ 
 at the time of birth against $a_{\rm birth}$, the cosmological scale factor at the time of birth. We also display the corresponding
 values for birth redshift $z_{\rm birth}$. The colour coding visualizes the density of stellar mass per pixel
 $n\left[log\left(R / R_{200}\right), a_{\rm birth}\right]/n_{\rm max}$ normalized to the maximum density $n_{\rm max}$
 of a specific panel.  Displayed models are 1646 (top left),  1196 (top right), 1192 (bottom left)  and AqC (bottom right). 
 For each displayed model, we give the fraction of accreted stars $f_{\rm acc}$ in the top right of the 
 corresponding panel. The vertical black lines indicate a radius of $0.14 \; R_{200}$, which we use to distinguish between
 in-situ and ex-situ star formation.}
\label{insi}
\end{figure*}

It is interesting to see that all of the higher and lower mass models are more similar in evolution to 1192 than to AqC, as they
show continuous mass growth at all radii. The detailed extent and distribution of mass growth varies significantly, but
the majority of these galaxies show more complex evolution than simple inside-out growth.
All the lower and higher mass galaxies are part of the halo sample first studied by \citet{oser}. The remaining ten haloes
are in equal parts made up of haloes from \citet{oser} and of Aquarius haloes \citep{aquarius, cs09}.
For these ten haloes A13 found SFHs in good agreement with abundance matching predictions by \citet{moster},
as well as $z=0$ masses which are similar to the ones of the galaxies studied by D13 and P13.
The Aquarius haloes
were selected to be prime candidates for hosting disc galaxies and have rather quiescent merger histories.
Four out of five of these (AqB is the exception) host galaxies with kinematic disc fractions above $50 \%$, higher
than any of the disc fractions found in non-Aquarius haloes. The structural evolution of all Aquarius haloes shows
an inside-out behaviour with varying details. Interestingly, the haloes from \citet{oser} that are closest to simple
inside-out formation are also in this mass range (the two galaxies in halo 0977, which are however about to merge after $z=0$).

\citet{perez13} recently presented an archaeological study of the 
growth of observed galaxies as a function of time and radius. They found
that the inferred relative stellar mass growth histories for their lowest
mass galaxies are similar at all radii with centres being tendentiously
younger than outskirts, whereas in the most massive galaxies the 
centres have grown significantly earlier than the outskirts. If we 
exclude our most massive galaxies, which overproduce stars at late times,
the remaining models cover a similar range in stellar mass as the 
observational sample. The corresponding panels in Fig. \ref{radial-all}
indicate similar trends as in observations. Splitting these model 
galaxies into three mass bins, we also find that the lowest mass 
galaxies show very similar growth rates for inner and outer regions. 
With increasing mass the difference between outer and inner growth rates
increases with inner parts being significantly older. We do, however,
not recover the trend of low mass galaxies having younger inner than 
outer regions.

In each panel of Fig. \ref{radial-all}, we give the Sersic index $n_s$ we find when fitting one-component Sersic profiles
to the $z=0$ surface mass density profiles. In A13, we have shown that most profiles show a down-bending break at outer radii.
We attempted to fit the profiles out to the break radii. Not all profiles are well fit by one-component profiles.
The resulting indices $n_s$ scatter between 1 and 4.
Interestingly, we find no correlation between $n_s$ and the kinetic disc fraction as determined in A13.
The four galaxies with highest kinematic disc fractions have $n_s$ values between 1.5 and 4. In AqC ($n_s=3.8$), the high
value results from the ring-like stellar disc structure around the central dispersion-dominated regions. There is a trend
of increasing Sersic index with galaxy mass, but not as clear as in real galaxies \citep{dutton09}.

In summary, we have shown that simulated disc galaxies exhibit a variety in morphologies and formation histories when analyzed 
with mock three-colour images, or through the evolution of surface brightness/density profiles. For the latter,
we find evolution ranging from inside-out growth of a disc around a central stellar population
which hardly evolves after $z=2$ to continuous mass growth at all radii at all times. 
We find more extended discs in optical light than in mass, as in observations, and
we identify complications in the interpretation of mass evolution
from light profiles especially for edge-on galaxies and for galaxies that grow mainly inside-out.

\section{Mergers and misaligned infall}

In our introduction we listed several mechanisms that lead to central mass growth in disc galaxies.
In A13 we already noted that our models contain only a small number of bars. Fig. \ref{evolution1} confirmed this and also
showed no examples of unstable, clumpy discs. This leaves mergers and accretion of gas with misaligned angular momentum
as possible mechanisms. We have shown that haloes selected to have more quiescent merger
histories show less central mass growth. In A13, we already showed that misaligned infall of gas can 
lead to the formation of a counter-rotating disc on top of an existing disc. 
In this section we study in more detail, how mergers and misaligned 
infall influence the structural history of our model disc galaxies.
Reorientation of disc galaxies by large angles in these simulations is always connected to misaligned infall,
which is why we do not analyze it separately here.

Motivated by \citet{oser}, we plot in Fig. \ref{insi} the distribution of stars in the plane of formation scale-factor
$a_{\rm birth}$ versus formation radius in units of the virial radius of the main galaxy $R / R_{200} (a_{\rm form})$.
Unlike \citet{oser}, we use 14 \% of the virial radius instead of 10 \% to distinguish between in-situ and ex-situ SF, as otherwise
the outskirts of several of our galaxies would be ex-situ. These plots show when stellar mass was accreted in mergers, as long as 
the merging galaxy is gas-rich and star-forming, which is the case for all the significant merger events in our simulations.

For the four galaxies shown, we give in the upper right of each panel the fraction of stars that were accreted, i.e.
which had birth radii $R_{\rm birth}>0.14\;R_{200,\rm birth}$. The fractions are between 8 \% and 18 \%, in agreement with the
finding of \citet{moster} that galaxies of these masses should be dominated by stars formed in-situ. The values are typical
for all our 19 model galaxies, which have accreted fractions ranging from 2 \% to 40 \%.

Comparing the panels for models 1192 and AqC, which we already studied in Fig. \ref{radial-g}, again reveals
striking differences. In AqC stars only form in the centre of the galaxy at $a<0.35$, when the galaxy undergoes several merger events.
Afterwards, in the absence of mergers, stars form only in a ring around the centre.
Merging activity continues until $a\sim 0.7$ in 1192 and continuously triggers central SF. When merging has ended,
there is a clear trend towards less centrally concentrated SF, despite the presence of a bar.
The mergers with the strongest impact have baryonic mass ratios of $\sim 1:2.5$ ($a\sim 0.4$) and $\sim 1:5.5$ ($a\sim 0.67$).
There are several more mergers with mass fractions of $\sim 1:10$ and smaller. The total accreted stellar mass is small (11.5 \%)
as all incoming galaxies have gas fractions between 50 \% and 90 \%.

The panel for model 1196 in Fig. \ref{radial-all} (see also sixth row in Fig. \ref{evolution1}) reveals inside-out growth 
at $z>0.5$, followed by an event of strong central growth, followed by inside-out growth at $z<0.3$. In Fig. \ref{insi}
we detect that a merger at $a\sim 0.65$ is the event which shapes the formation history. A closer analysis shows that the merger
has a baryonic mass ratio of $\sim 1:3$ and that the involved galaxies both are gas-rich with gas fractions $M_{\rm cold gas}/M_{baryons}(<20\;{\rm kpc})$
of 43 \% and 55 \%.

\begin{figure*}
\centering
\vspace{-1cm}
\includegraphics[width=17cm]{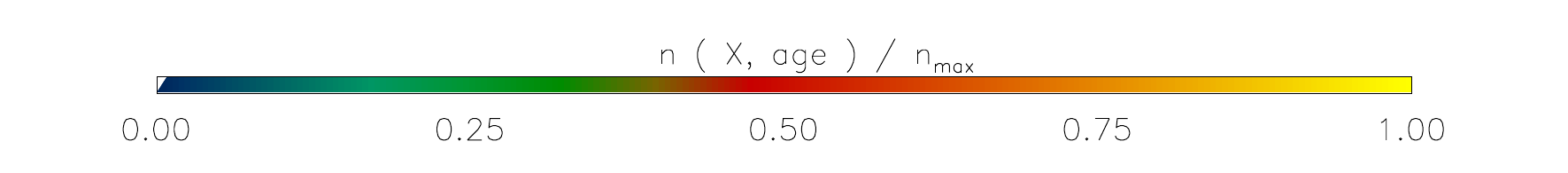}\\
\vspace{-0.25cm}
\includegraphics[width=8.cm]{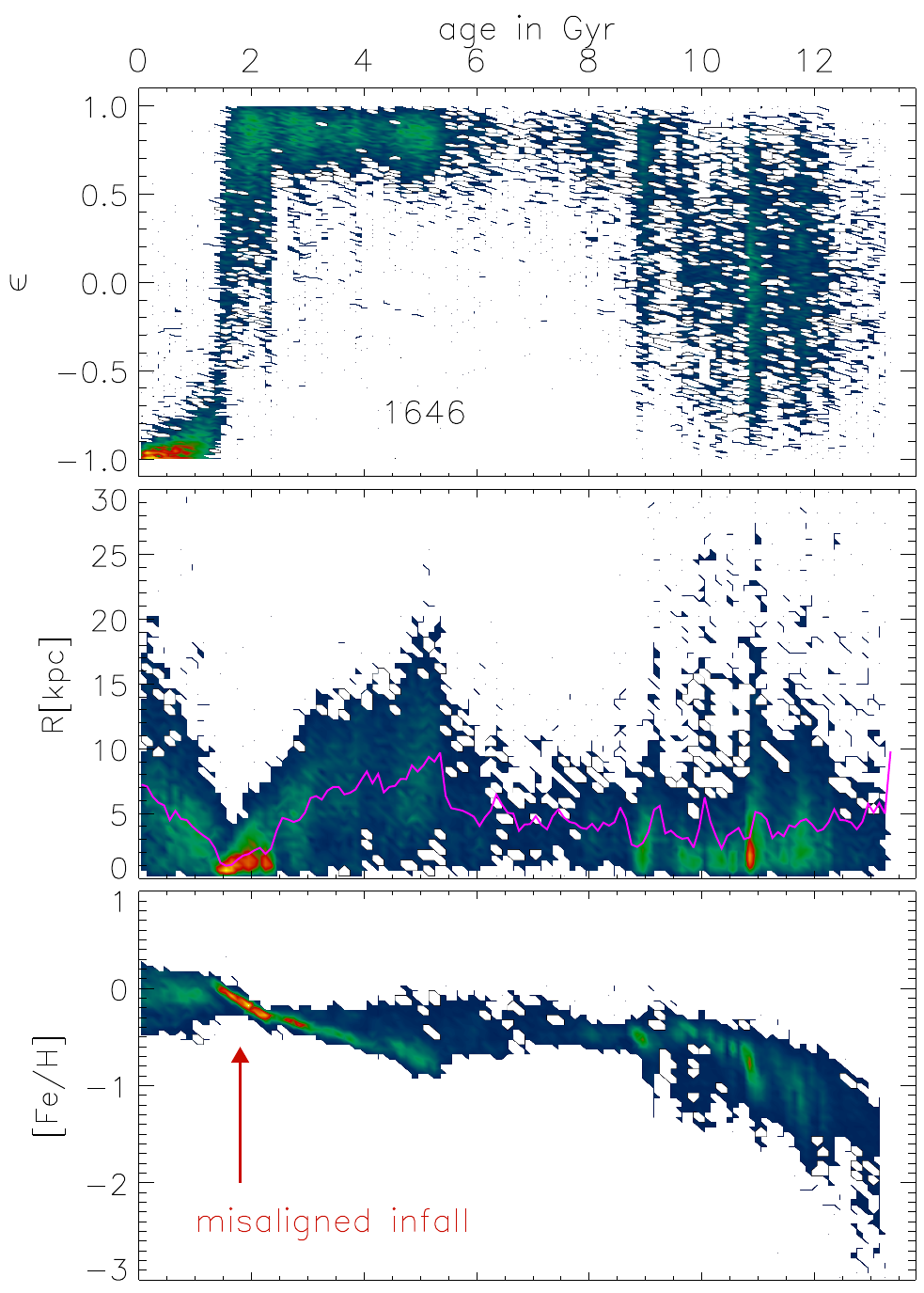}\includegraphics[width=8.cm]{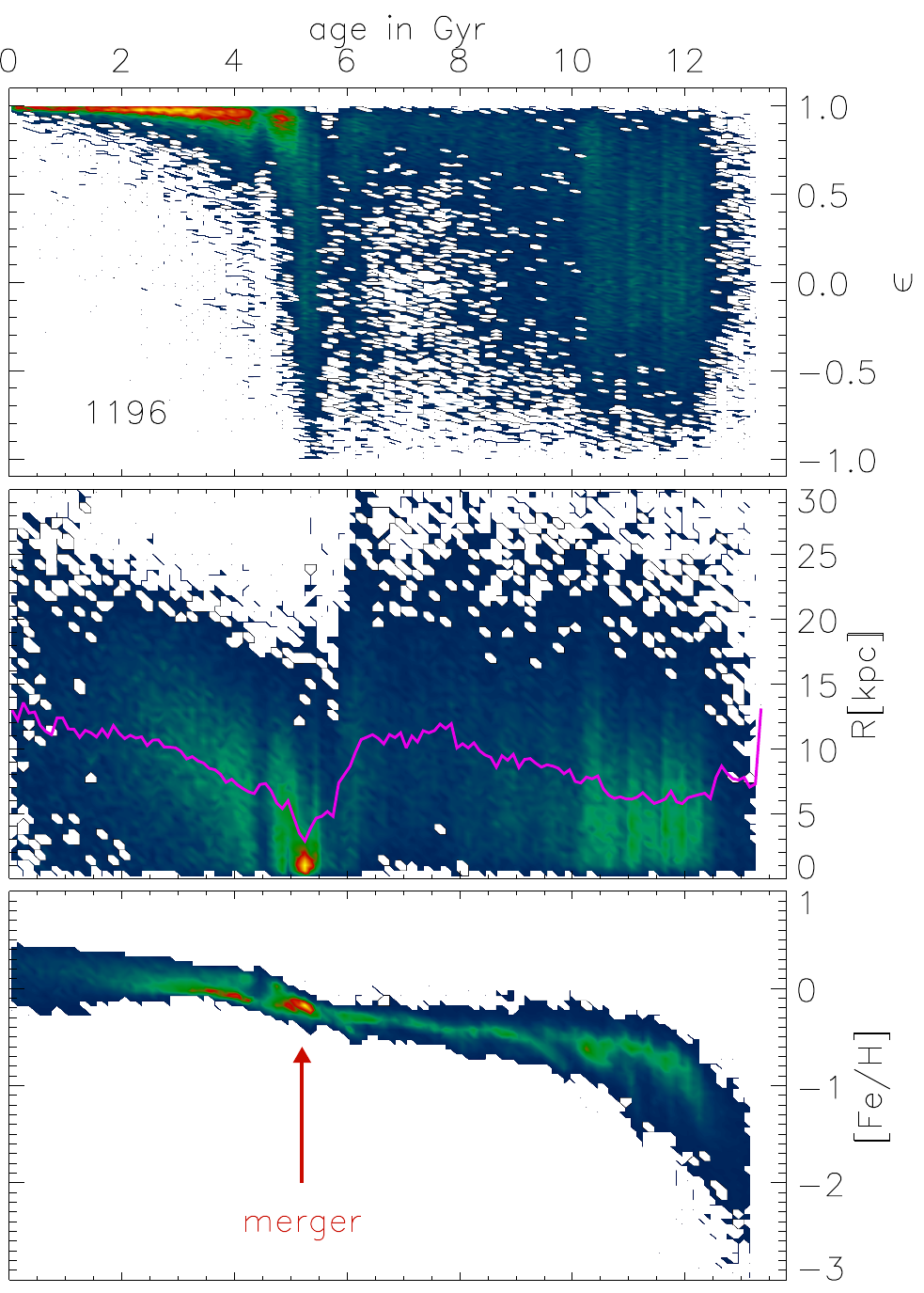}\\
\includegraphics[width=8.cm]{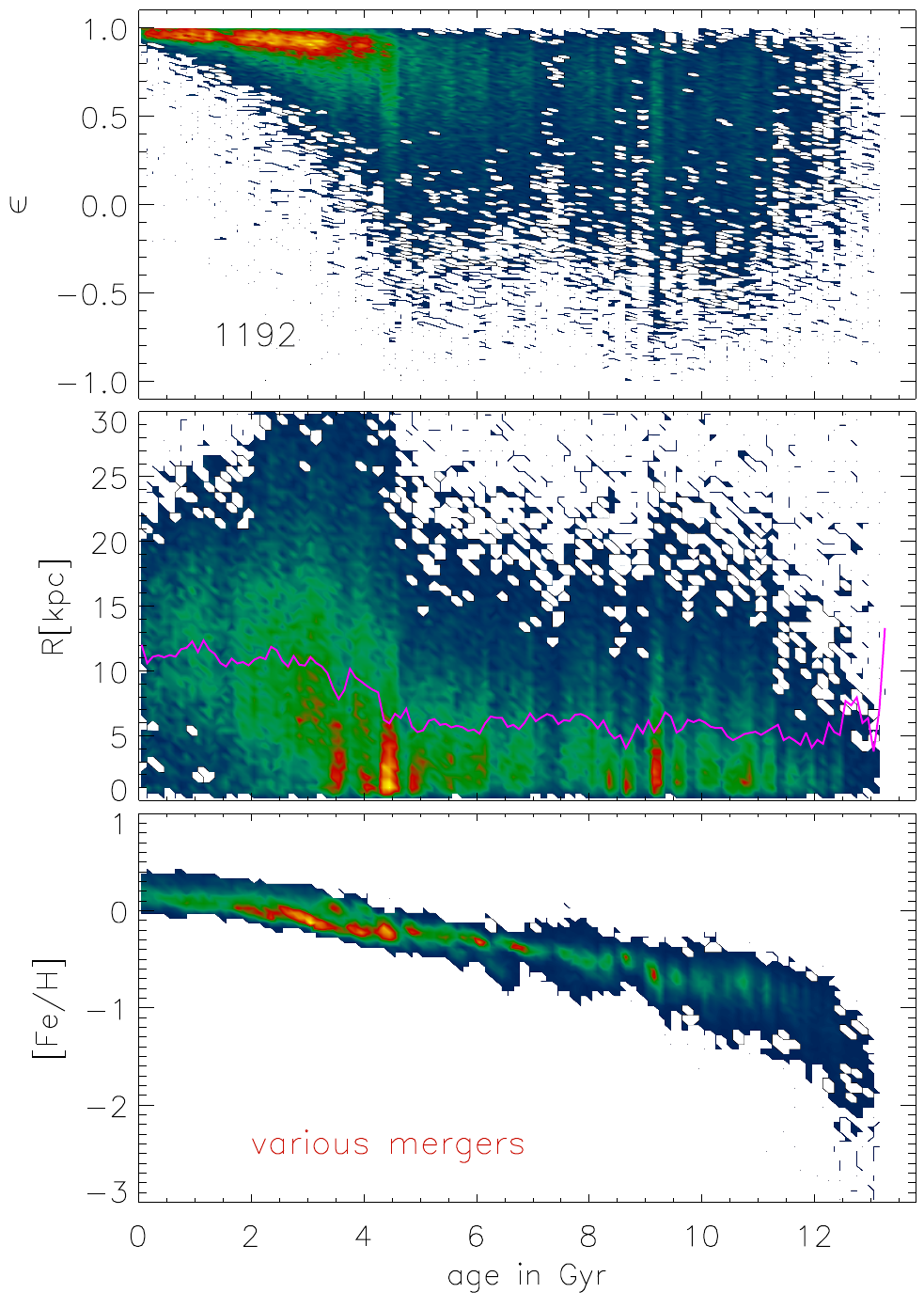}\includegraphics[width=8.cm]{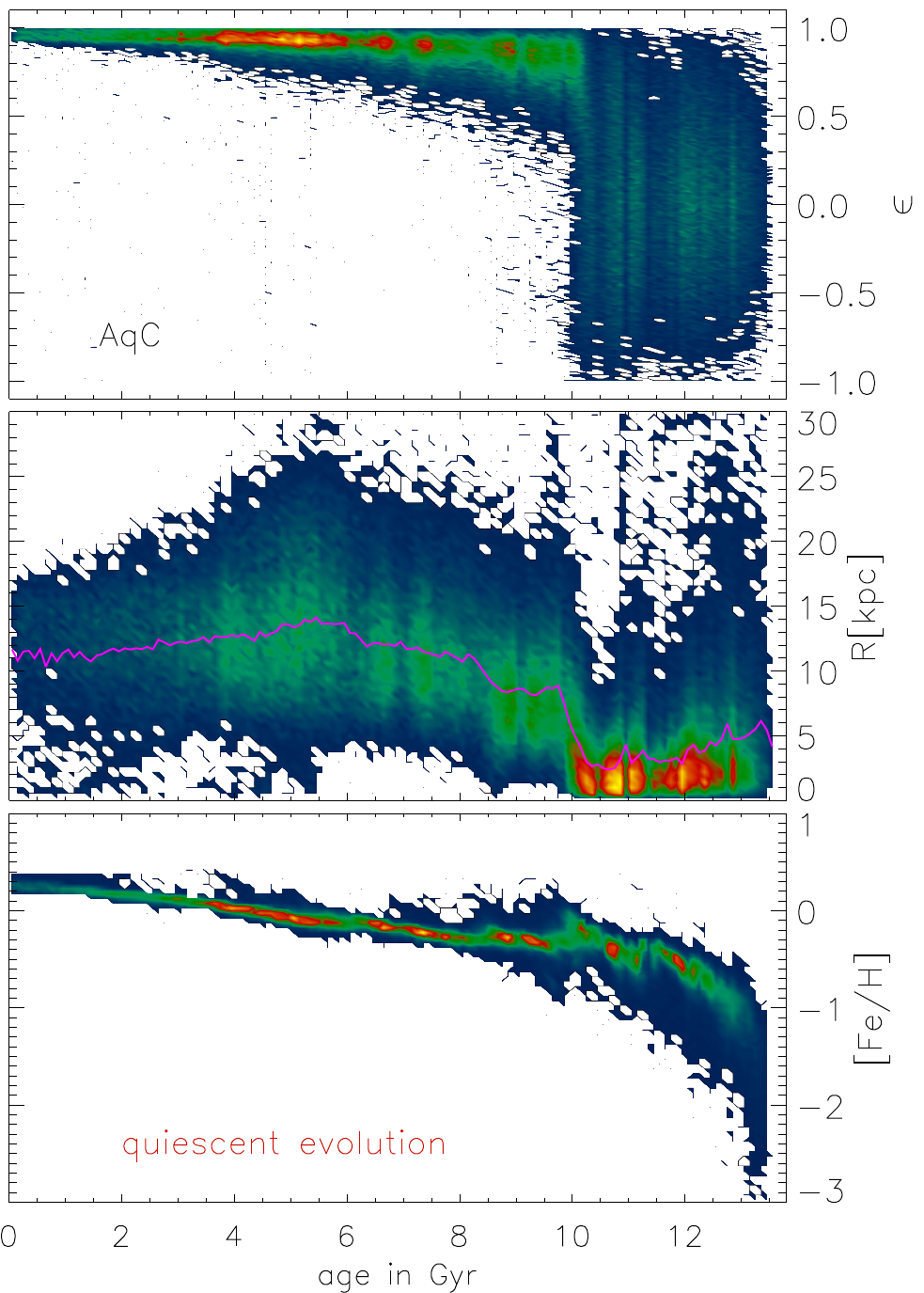}\\
\vspace{-0.25cm}
\caption[Archaeological structural quantities]
{Distributions of $z=0$ galactic stars in models 1646 (top left), 1196 (top right), 1192 (bottom left) and AqC (bottom right)
in the planes of $z=0$ circularity $\epsilon$ (top), cylindrical radius $R$ (middle) and metallicity $[Fe/H]$ (bottom) against age.
The magenta lines in the middle panels indicate the mean $z=0$ radius of stars of a certain age.
The colour coding visualizes the density of stellar mass per pixel  $n\left({\rm X,\, age} \right)/n_{\rm max}$ (where X is
$\epsilon$, $R$ or$[Fe/H]$) normalized to the maximum density $n_{\rm max}$ of a specific panel.}
\label{archae}
\end{figure*}

Model 1646 (see also third row in Fig. \ref{evolution1}) 
does not undergo significant mergers after $a=0.4$. However at $a\sim 0.9$ an event with similar effects
on the central mass growth as displayed by the merger in model 1196 is clearly visible. The corresponding panel in Fig. \ref{radial-all}
also clearly shows this event. In A13 we found that the youngest stars of this model lived in a counter-rotating disc, resulting from
misaligned infall of gas, which leads to the shrinking and reorientation of the gas disc, to an episode of centrally
concentrated SF and to inside-out growth of the counter-rotating disc.

In Fig. \ref{archae} we take an archaeological approach to these questions and analyze at $z=0$ the radii, 
the circularities, $\epsilon=j_{\rm z}/j_{\rm circ}(E)$, and the metallicities of stars as a function of their age. 
When comparing these radii to the birth radii displayed in Fig. \ref{insi}, keep in mind that especially, but not exclusively, the accreted 
stars have significantly different radii at $z=0$ than at birth. Note that the upturns for the $z=0$ radii of the oldest
stars are caused by accreted stars.

We start with a combined analysis of radii and circularities. In AqC all stars with ages $\tau \lesssim 10$ Gyr live in a disc,
as all these stars have circularities $\epsilon>0.75$. The circularities of older stars are centered around $\epsilon=0$ and their radii are
centrally concentrated, they form a bulge. The bulge population indicates several episodes of increased SF activity, which are connected to 
the various merger events visible in Fig. \ref{insi}. After the last destructive merger, the disc formed inside-out
($6<\tau/{\rm Gyr}<10$), then SF started to decline and the disc stopped growing in size.

The situation in model 1196 is complicated by the additional late merger event at $\tau\sim 5$ Gyr, which triggers a short 
episode of centrally concentrated SF resulting in a dispersion-dominated component. The old bulge component with ages $\tau>10$ Gyr 
is more extended than in AqC, in part due to the higher fraction of accreted stars. Between early bulge formation and the late merger
($5<\tau/{\rm Gyr}<10$), a thick, but rotationally supported component is present, which forms inside-out.
The $z=0$ thin disc component also formed inside-out over the last 5 Gyr.

As in 1196, model 1646 shows an old bulge and an intermediate-age thick disc component, which formed inside-out.
However, the infall of misaligned material, which started $\sim 5$ Gyr ago, led to a shrinking of the gas disc
which first created a centrally-concentrated, non-rotating component and then an inside-out growing, counter-rotating disc.

The various minor mergers of model 1192 lead to a multitude of episodes of centrally-concentrated SF up until $\sim 4$ Gyr 
ago. These stellar populations are, however, mildly rotating at $z=0$, with a mean circularity $\epsilon\sim 0.5$. At late times,
the mean stellar radius increases only mildly, as the disc is barred, producing a broader $\epsilon$ distribution
compared to the other young discs.

In Fig. \ref{archae} we also plot age-metallicity relations (AMRs) for all stars in the galaxies. Note that in simulated 
galaxies, this relation depends strongly on the details of the modelling of metal enrichment and diffusion, as was e.g. shown  by 
\citet{pilkington}. For comparison, the solar neighbourhood shows a very flat AMR for stars with ages $\tau \lesssim 10-12 $ Gyr
and a steep decrease for older stars \citep{casagrande}.

All four depicted galaxies show a steep decrease in $[Fe/H]$ at $\tau \sim 12$ Gyr. This corresponds to the first stages of galaxy formation
when the first generations of stars enriched the ISM to metallicities $[Fe/H]>-1$ . The scatter at these ages is large, as
the accreted fractions are high and the different progenitor galaxies show variations in enrichment history. As for the solar neighbourhood,
there is a transition in the AMR of all four galaxies at $\tau\sim10-12$ Gyr. The detailed evolution for younger stars differs between
model galaxies.

We note that the phases of inside-out growth identified above are characterized by rather flat AMRs, e.g. at $5<\tau/{\rm Gyr}<10$ in 1196
or at $3<\tau/{\rm Gyr}<9$ in 1646. The merger or misaligned infall events in these models show short phases of increasing $[Fe/H]$.
In AqC the metallicity is mildly increasing with time, more strongly at $\tau<6$ Gyr, when the disc no longer increases its size and enrichment
by stars dominates over accretion of less-enriched material. The numerous minor mergers in 1192 lead to a continuously increasing AMR
which tends to flatten at late times after the last merger. In 1646, at the start of the outside-in evolution ($\tau \sim 5$ Gyr),
the average metallicity drops and the spread in $[Fe/H]$ widens, as new, relatively metal-poor gas is accreted after a gas-poor phase
with little SF.

If we, instead of all galactic stars, only consider the AMR of outer disc stars ($R>5$ kpc), we can give a tentative statement on whether
one of the metallicity features found for our models could also be found in a solar-neighbourhood-like environment. We find that the details
of the AMR vary, but the general shape is the same in the centre and in the outer regions.

\begin{figure}
\centering
\includegraphics[width=9.cm]{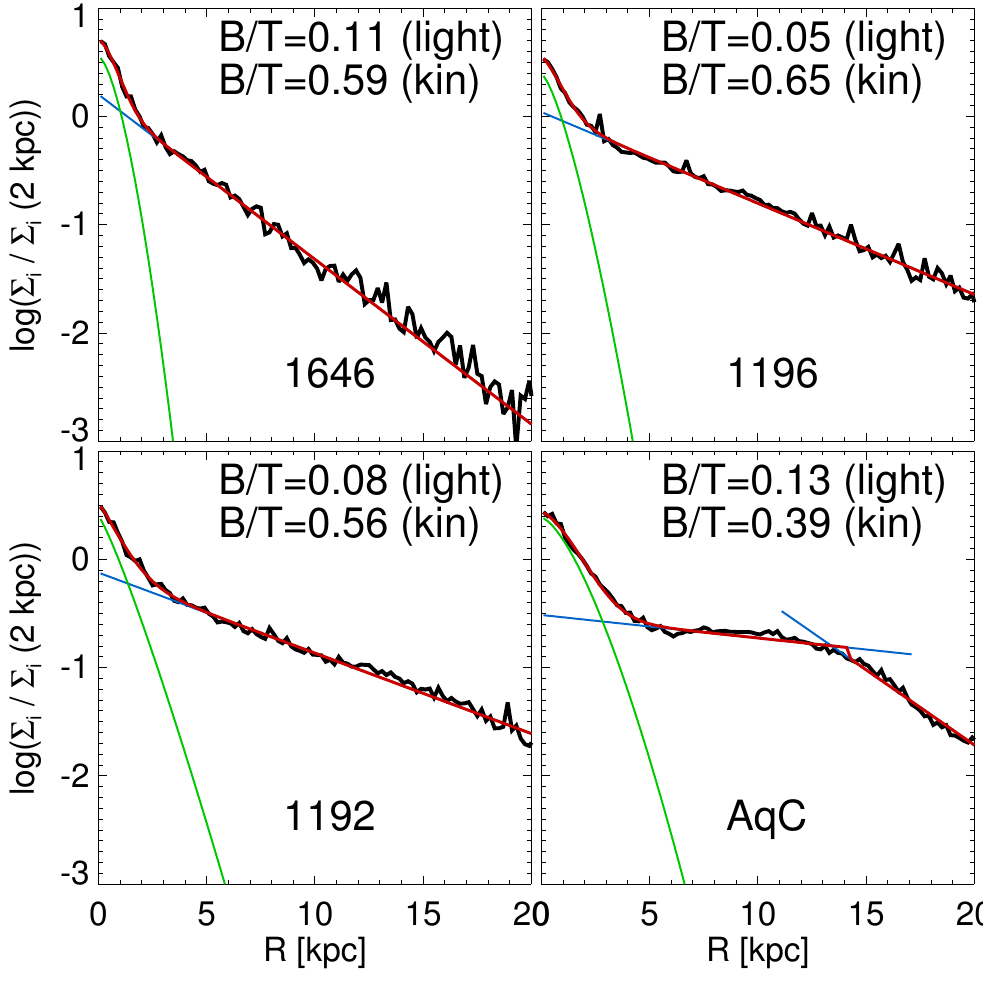}
\caption[I band profile fits]
{Face-on $z=0$ mock $i$ band light profiles, normalized to the value at $R=2\;{\rm kpc}$ for models 1646, 1196, 1192, AqC.
We overplot the result of exponential disc $+$ Sersic bulge fits to the profiles. Red is the complete fit, green the
Sersic profile and blue the exponential. In case of AqC we fit the disc part with two exponentials and a down-bending break.}
\label{prof}
\end{figure}

It is interesting to ask how these different formation histories connect to bulge-to-total ratios $B/T$. In A13 we determined
kinematic disc and bulge ratios by isolating disc regions in the distribution of stars in the plane of
circularity $\epsilon$ vs. age $\tau$ (see Fig. \ref{archae}). 
Disc regions are defined by connected areas in these diagrams, where stars with $\epsilon>0.7$ are dominant, e.g.
at $\tau<10.5 {\rm Gyr}$ for model AqC. For 1646, we here add up contributions from the co- and the counter-rotating disc.

For observations, $B/T$ is often determined by fitting a profile consisting of an outer exponential disc profile and an inner
bulge Sersic profile to surface brightness profiles. The contribution from the exponential is then counted as the disc component
and the inner Sersic profile as the bulge component. For our models, we have applied this method to face-on mock $i$-band profiles, which were 
created as described in Section 3. Four fits for the models 1646, 1196, 1192 and AqC are displayed in Fig. \ref{prof}.
For AqC and similar models, which show a down-bending break, we fit the disc part with two exponentials.
For the inner Sersic profiles we typically find indices $n_S\sim 1$, corresponding to pseudo-bulges.
We have performed these bulge/disc decompositions at $z=0$ except for the haloes 0977 and 2283,
where we go back to $z\sim 0.1$, as both haloes host two interacting galaxies at $z=0$. 

In Fig. \ref{prof} we display kinematic and mock-photometric $B/T$ values for the four discussed models. All models show significantly lower
values of $B/T$ in the photometric decomposition than in the kinematic method. This conclusion has previously been drawn by \citet{cs10}
and \citet{marinacci}. AqC has the lowest kinematic bulge fraction, as there are no destructive events after $z\sim 2$ (see Fig. \ref{archae}). However,
its photometric $B/T=0.13$ is higher than that of the other three displayed models ($B/T\le0.11$), which do experience destructive events after $z=1$
and consequently have kinematic disc fractions below 50 per cent.
In AqC, there is hardly any SF in the centre after $z=2$ (see Fig. \ref{radial-g}).
The central profile of the old bulge component at $R<5\;{\rm kpc}$ is distinct from the disc part, 
which due to its ring-like nature, is fit by a flat part, a down-bending break and an exponential outer part.
For the other cases, there has been more central SF activity at later times and the bulge excesses are less distinct.

\begin{figure}
\centering
\includegraphics[width=8.5cm]{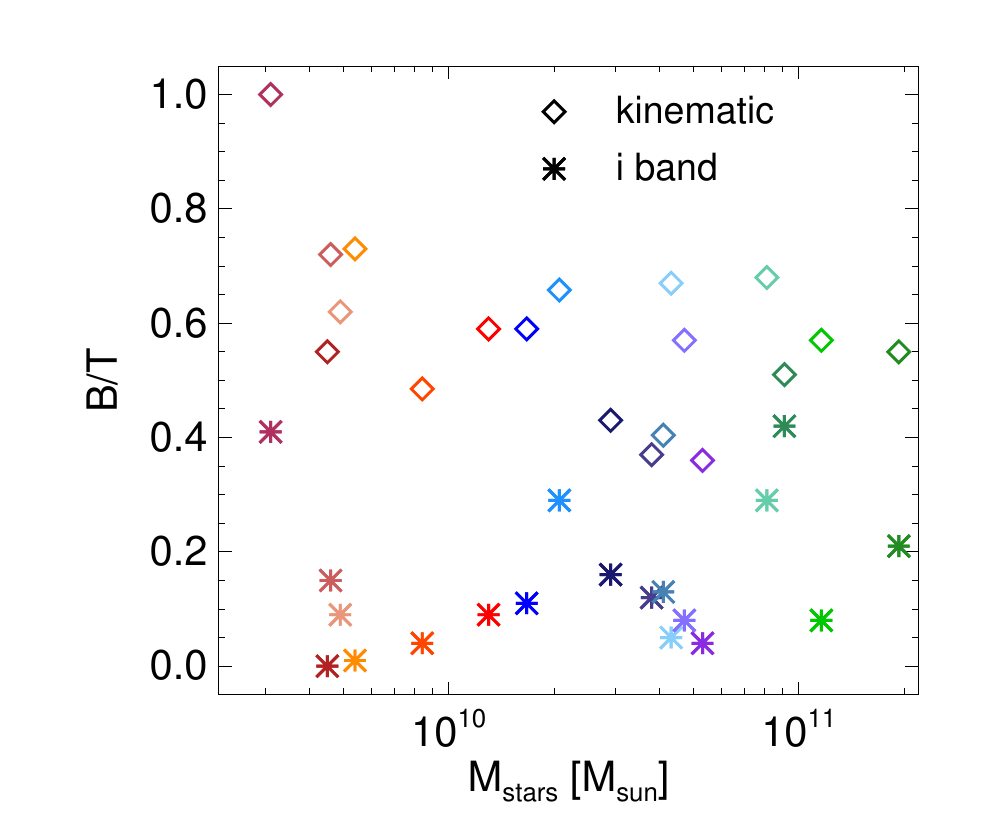}
\caption[Bulge Fractions]
{Bulge-to-total ratios $B/T$ as a function of stellar mass $M_{\rm stars}$ for all 19 galaxies. For the galaxies in the two haloes
hosting two galaxies, galaxies are shown at $z=0.1$, the rest is at $z=0$. We show kinematic ratios following the method described 
in A13 (diamonds) and ratios determined from $i$-band light profiles by fitting an exponential disc $+$ Sersic bulge profile (stars).}
\label{bulges}
\end{figure}

In Fig. \ref{bulges}, we show $B/T$ values for all 19 galaxies as a function of stellar masses $M_{\rm stellar}$.
For kinematic bulge fractions, there is a trend of decreasing $B/T$ with increasing mass, which is, however, driven by the four
truly disc dominated galaxies. These all live in Aquarius haloes, which have a limited mass range and a quiescent merger history.
Moreover, considering that the lowest mass models are older
than observed counterparts, whereas the highest mass models form too many stars at late times, the trend might invert, if
SFHs were reproduced correctly at late times. Generally, there is no clear correlation visible between kinematic and photometric
$B/T$. There is only one galaxy which is both among the five models with highest kinematic disc fractions and the five models
with highest photometric disc fractions. 

\begin{figure*}
\centering
\includegraphics[width=18cm]{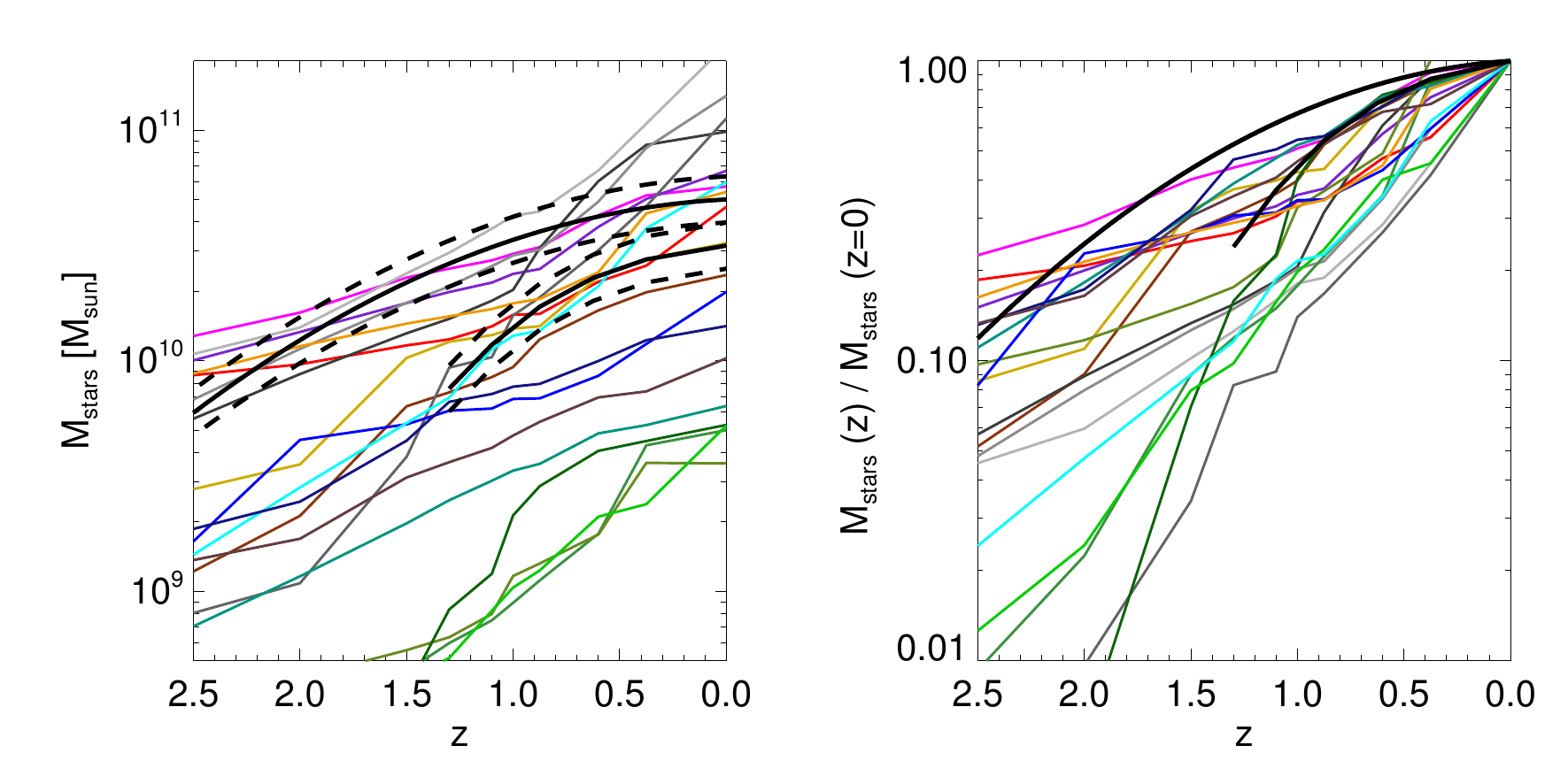}
\caption[The evolution of simulated galaxies in stellar mass]
{The evolution of stellar mass $M_{\rm stars}$ with redshift $z$ for all 19 model galaxies.
 Left: The evolution of absolute masses $M_{\rm stars}(z)$. Right: The evolution of masses in units of the $z=0$ mass
 $M_{\rm stars} (z) / M_{\rm stars} (z=0)$ for each galaxy.
 We overplot in black the mean mass evolution histories adopted by D13 (final mass $5\times 10^{10} M_{\odot}$) and P13 
(final mass $3.2\times 10^{10} M_{\odot}$).  Dashed lines represent a scatter of $\pm 0.1 {\rm dex}$.}
\label{massevo}
\end{figure*}

Interestingly, there is a trend of increasing photometric $B/T$ with stellar mass, as for observed galaxies (e.g. \citealp{dutton09}).
Several of our models have very low photometric $B/T<0.1$ and some of them thus might correspond to observed `bulgeless' galaxies
(e.g. \citealp{kautsch}). As in observations \citep{fd11}, a high fraction of model galaxies with stellar masses
$M_{\rm stars}<3\times10^{10}M_{\odot}$ shows pseudobulges with $B/T<0.1$.
These results suggest that, unlike often assumed, observed galaxies with very high photometric disc fractions
might not all have had very quiescent formation histories.

Clearly, photometric decompositions depend on the applied wavelength range and there are different approaches to determining 
kinematic disc fractions (see A13). We have checked how other methods would affect our results and found that our conclusions 
are robust against the details of the photometric and kinematic decomposition methods.

In summary, we have shown that, for our simulated disc galaxies, gas-rich mergers and misaligned infall events are the main drivers of central mass growth.
Galaxies with high $z=0$ kinematic disc fractions, which generally show no such event after $z\gtrsim 1.5$, hardly grow in central mass.
For other galaxies, such events transport low angular momentum gas inwards leading to in-situ formation
of dispersion-dominated stellar populations. `Archaeological' investigations of stellar radii, circularities
and metallicities at $z=0$ reveal these events as centrally-concentrated, non-rotating populations of stars of a small age range,
during which metallicity increases more strongly than average. An investigation of kinematic and photometric disc fractions
has shown that photometric values for $B/T$ are always lower than kinematic ones and correlate little with those.
Galaxies with low photometric bulge fractions ($B/T<0.1$) can harbour more than 50 per cent of their stars in non-disc components,
especially in our lower mass models.

\section{Comparison to observations}

\begin{figure*}
\centering
\includegraphics[width=16cm]{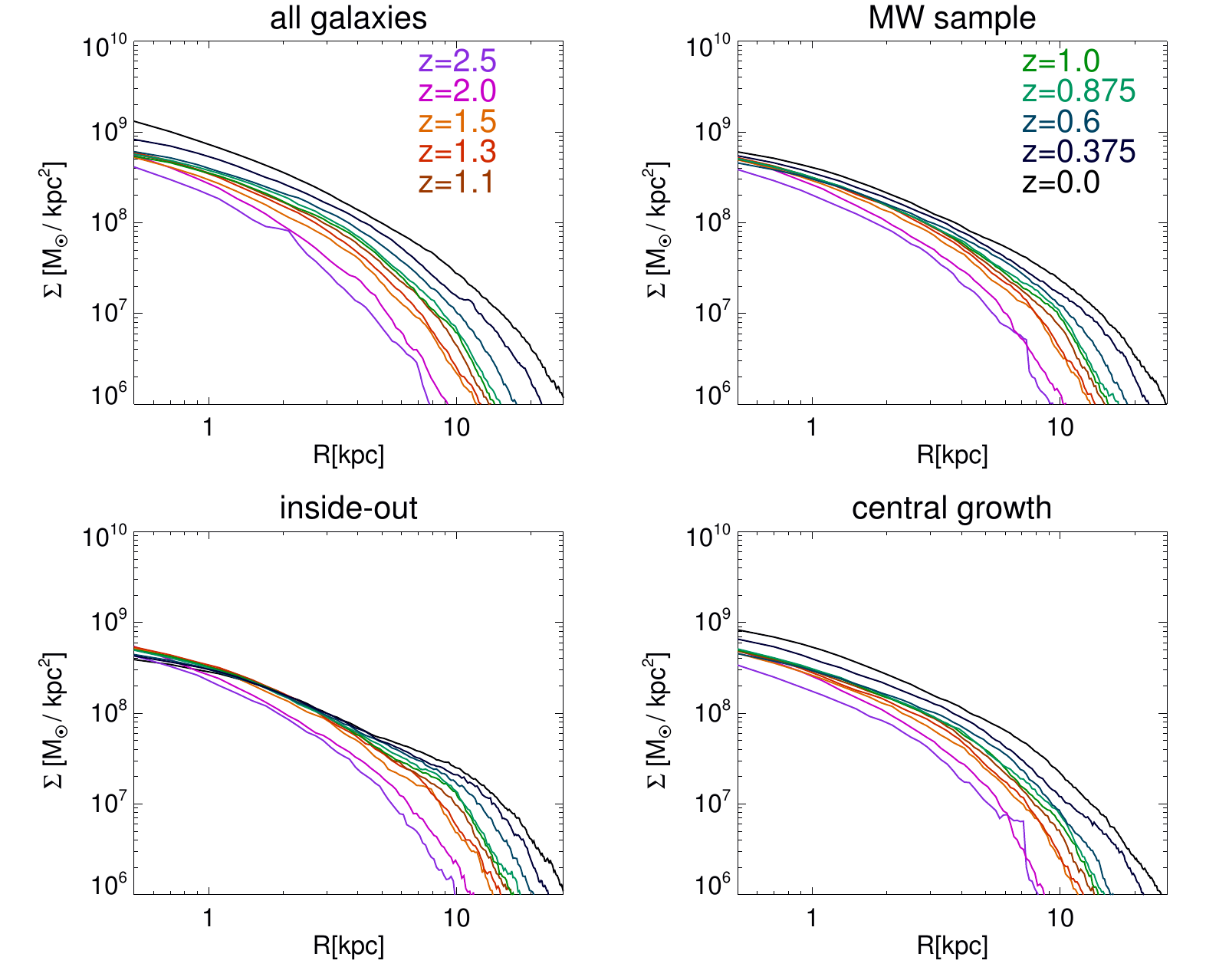}
\caption[Means mass and light profile evolution for various samples]
{Evolution of mean number-weighted stellar mass surface density profiles from $z=2.5$ to $z=0$ for various samples:
 all galaxies (upper left), MW mass sample (upper right), inside-out sample (lower left), central growth sample (lower right).
 We here apply scaling in mass and radius according to the procedure detailed
 in the text. Masses are scaled to the mean mass evolution of the simulated samples.}
\label{1means}
\end{figure*}

In this section we attempt a more direct comparison of our simulations to the observations by D13 and P13.
This is complicated by the variety in mass evolution histories of our galaxies, few of which agree with the
selection criteria of either study. It is also unclear, how well the selection criteria
used by P13 and D13 actually trace the mean evolution of galaxies at a targeted $z=0$ mass.
\citet{leja} showed that the method applied by D13 worked reasonably well in tracing galaxy populations
in semi-analytical models of galaxy formation. P13 showed that their selection criteria agree
with predictions for galaxy populations from abundance matching.
In Fig. \ref{massevo} we plot the evolution of stellar mass of the model galaxies against redshift in two ways:
in the left panel we show $M_{\rm stars}(z)$ to depict the variety among masses of the model galaxies at each redshift
and in the right panel we show $M_{\rm stars}(z)/M_{\rm stars}(z=0)$ to highlight the variety in stellar mass assembly history.
In both panels we overplot the corresponding lines for the selection criteria of D13 and P13.

The range of stellar masses of our model sample is at all times much wider than the selection intervals.
Moreover, most of the galaxies that at some point in time are inside one of the intervals, fulfill this criterion
only temporarily. In other words, most individual formation histories are flatter, steeper or more episodic than the smooth
average formation history of a population selected by a certain $z=0$ mass. Galaxy 0858, for example, at $z=2.5$
is an order of magnitude less massive than the mass anticipated by D13, but at $z=0$ more massive
than their observed galaxies. The right panel of Fig. \ref{massevo} reveals that none of our galaxies lies close
to the relative assembly history selected by D13, but several galaxies are similar to the history assumed by P13.
As was shown in Fig. 3 of A13, the expected peak in SFH at $z\sim1-2$ of 
galaxies with $z=0$ stellar masses similar to that of the MW (e.g. \citealp{moster})
is not well reproduced by our models. The models on average show a flatter SFH(z), which contributes to the discrepancy
in the right panel of Fig. \ref{massevo}. 

Moreover, as already noted in A13 and in Section 3, the four most massive simulated galaxies show overly high SFRs
at $z<0.5$ compared to observed galaxies of similar mass. If a quenching mechanism, such as AGN feedback, were present
in the simulations at late times, the shape of their SFHs would likely be more similar to the assumptions of D13.
Considering the lower mass models, their relative mass assembly histories are on average steeper than the ones of
higher mass galaxies. This is a hint of `downsizing' in galaxy evolution (e.g. \citealp{fontanot}), as was recently also 
reported for the simulations of \citet{derossi}. However, as we have shown in A13, the trend in our simulations is not
strong enough, as the average ages of our low mass galaxies are significantly higher than those inferred for observed galaxies.

Nevertheless, we create four test samples of galaxies to be compared to the observational results. The first sample includes all
galaxies, the second sample excludes the four most massive and five least massive galaxies at $z=0$.
This leaves ten galaxies with $z=0$ stellar masses $1\times 10^{10}<M_{\rm stars}/M_{\odot}<6\times 10^{10}$, which we sloppily call
`MW mass sample'. Due to the variety of formation histories of our galaxies, we further divide the 10 galaxies in equal parts into an `inside-out'
and a `central growth' sample according to the growth of stellar mass at $R<2$ kpc at $z<2$.
We note that our general results are not affected by our selection criteria.
Selecting at each $z$ the galaxies closest to the mass evolution assumed by D13 would, for example, give results similar to that of the
MW mass sample.

\begin{figure*}
\centering
\includegraphics[width=18cm]{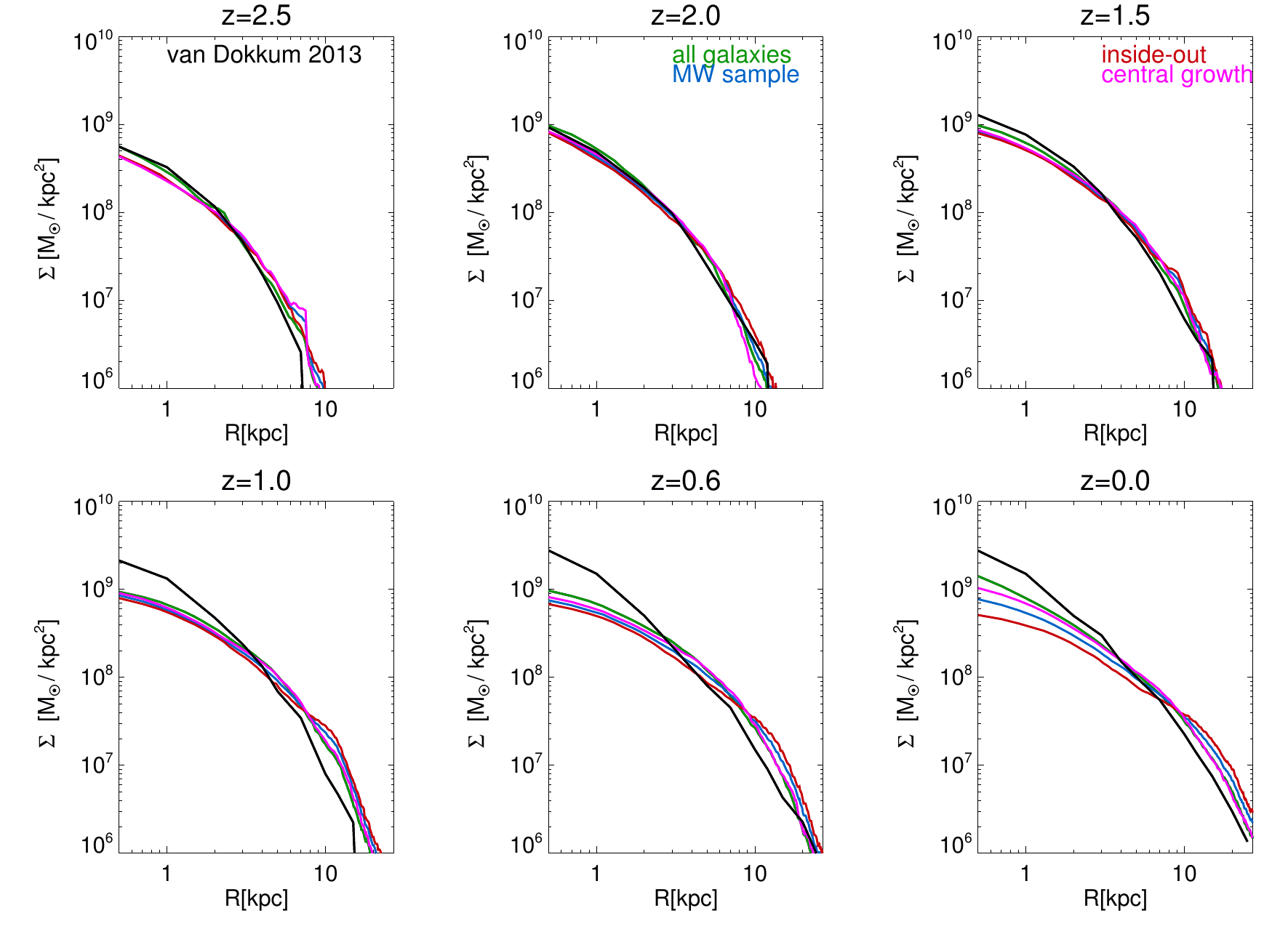}
\caption[Direct comparison of observed and simulated profiles]
{A direct comparison of simulated and observed mean profiles. In black we show the observed profiles by D13 at
 $z=2.5$, $2.0$, $1.5$, $1.0$, $0.6$ and $0.0$. We show mean number-weighted mass surface density profiles $\Sigma(R)$ for four sub-samples
 of our simulated galaxies: all galaxies (green), MW sample (blue), inside-out (red) and central growth (magenta).
 For our scaling in mass and radius we here assume the mass evolution selected in D13.}
\label{compro}
\end{figure*}

Due to the wide spread in masses of our models we apply a scaling in mass and size. In terms of mass, we consider two cases:
scaling the integrated total mass of each galaxy to the mean mass of the considered sample 
at the given $z$ or scaling the galaxy mass to the corresponding mass selected by D13.
In terms of size, \citet{shen03} give a mass-size relation for $z=0$. At high-$z$, a mass-size relation with increasing
sizes for increasing stellar masses is also present \citep{mosleh, nagy, ichikawa}, but the shape of the relation is yet uncertain.
\citet{ichikawa} suggested that the mean slope is independent of redshift.
We therefore assume that the mean relative difference in size of galaxies at varying masses at all redshifts
is given by the \citet{shen03} relation.
Consequently, we extend or shrink the mass distribution of each galaxy in consideration by a factor that is determined
by the ratio of the Shen relation radii at
its actual mass and the assumed mean mass. We note that the general conclusions of our work are not dependent on the
details of the scaling procedure.

The evolution of the mean surface density profiles for all our four samples is presented in Fig. \ref{1means}. For this figure, we
scale each galaxy to the mean sample mass as detailed above and then 
stack the profiles to determine the mean number-weighted profile of the sample.
Note that the mean masses at a given redshift differ significantly from sample to sample.
Whereas the evolution of the outer surface density profiles ($R\gtrsim5$ kpc) for all samples shows a continuous increase in mass
and also a continuous increase in size, the samples significantly differ in the evolution of the central surface density.

\begin{figure*}
\centering
\includegraphics[width=18cm]{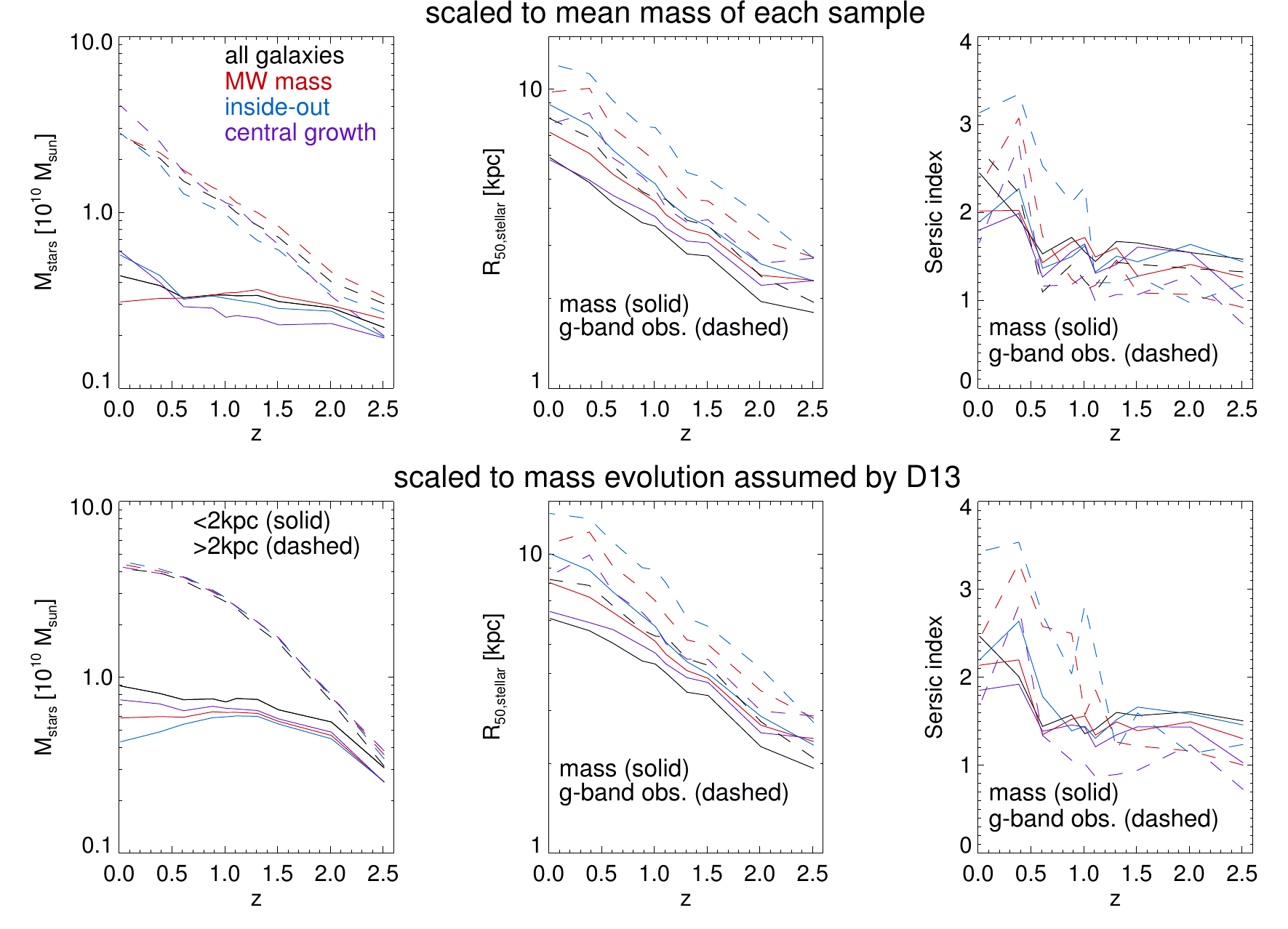}
\caption[The evolution of profile related quantities for various samples]
{Evolution of properties characterizing the mean mass and light profiles of simulated galaxy samples:
 All galaxies (black), MW mass (red), inside-out (cyan) and central growth (purple).
 Upper panels: Galaxies scaled to the mean mass evolution of each sample.
 Lower panels: Galaxies scaled to the mass evolution selected by D13.
 Left: Evolution of the mass within $R=$ 2 kpc (solid) and outside $R=$ 2 kpc (dashed). 
 Middle: Evolution of the half-mass (solid) and half $g$ band light (dashed) radii.
 Right: Evolution of Sersic indices obtained by fitting the mean mass (solid) and mean $g$ band light (dashed) profiles.}
\label{evols}
\end{figure*}

The `MW mass' sample shows significantly less central growth than the `all galaxies' sample. Fig. \ref{radial-all} reveals that
the four most massive galaxies drive central mass growth in the all galaxies sample at late times, whereas the low-mass galaxies are responsible
for the central growth at earlier times. By construction, the inside-out sample shows a decreasing mean central mass density with time
due to stellar mass loss as discussed in Fig. \ref{radial-g} for model AqC. Also by construction, the central growth sample shows higher
central growth rates than the MW mass sample.

In Fig. \ref{compro} we present a direct comparison of the samples with the observations of D13 at various redshifts.
As discussed above, the evolution of mean masses differs between samples and is also different from the evolution
assumed in observations, which is why we restrict ourselves to profile shapes and consequently scale all profiles
to the mass evolution assumed by D13.

We first note that the mean profile of all simulated galaxies is at all times more concentrated than the other three samples,
which is caused by lower and higher mass galaxies, which all do not show inside-out growth, as discussed in Section 3.
At $z\ge1.5$, the differences are, however, minor and the mean profiles of all samples show a remarkably good agreement with 
observations. Considering the error bars presented in D13, all samples are consistent with the data at these times.

After $z=1$ there is, however, a clear trend for all simulated galaxies to produce too shallow surface density profiles.
The trend is strongest for galaxies that grow inside-out with discrepancies increasing with time. 
At $z=0$, their scaled mean central surface density is a 
factor of $\sim 4$ lower than in observed galaxies and the scaled mean surface density at $R\sim 5-20$ kpc 
is consequently a factor of up to $\sim 3$ higher.
The sample of all galaxies at $z=0$ disagrees less strongly with the data, the scaled profile at
all radii agrees to within a factor of $\sim2$. The other two samples fall between all galaxies and the inside-out sample.
Note that the discrepancies would be smaller if no scaling in size were applied.
We caution that D13 used rest-frame $g$ band as a proxy for mass.
If we used mock $g$ band surface brightness profiles, the discrepancies at low $z$ would increase (see Fig. \ref{radial-g}).
For rest-frame $i$ band profiles, we find smaller discrepancies from mass profiles.

In Fig. \ref{evols} we show the evolution of several structural parameters characterizing the profiles, which were
also presented in D13 and P13. We show the quantities for both variants of scaling as explained above.
In the left panels of Fig. \ref{evols} we plot the evolution of the total stellar mass within and outside 2 kpc. 
D13 found an increase at $R<2$ kpc by a factor of $\sim 3.5$ from $z=2.5$ to 0.5 and no evolution afterwards, and a 
mass increase by a factor of $\sim 10$ outside.  P13, who analyzed slightly less massive galaxies, found a less flattened 
central mass increase at late times, which is likely connected to lower mass galaxies assembling stellar mass later 
(e.g. \citealp{moster}). They otherwise found similar results. Moreover, \citet{perez13} found that galaxies with stellar masses
below $10^{10}M_{\odot}$ have likely experienced very similar mass growth histories at all radii, whereas galaxies with MW-like masses
have grown inside-out. As discussed in Section 3, the latter result is qualitatively reproduced by our galaxy sample.

We first discuss results for the upper left panel, where we use the mean mass evolution of each sample for the scaling procedure.
When we consider all galaxies, we find an increase at $R>2$ kpc by a factor $f_{>}\sim 19$, significantly 
higher than observed. At $R<2$ kpc we find an increase by a factor of $f_{<}\sim3.9$, which agrees with observations. However, the
central mass growth does not flatten at late times. For the `MW mass' sample, the respective growth factors are
$f_{>}\sim 9$ and $f_{<}\sim2.0$, showing that these galaxies mainly differ from observations in the lack of
central growth. This discrepancy is consequently worse for the inside-out sample ($f_{<}\sim1.25$, $f_{>}\sim 8.3$) and rather small
for the central growth sample ($f_{<}\sim2.9$, $f_{>}\sim 10.5$).

As discussed above, the samples all show evolution in total stellar mass different from the galaxy samples in the observational papers.
In the lower left panel of Fig. \ref{evols} we therefore show the evolution of stellar mass within and outside 2 kpc when
we use the mass evolution assumed by D13 for the scaling procedure.
As already noted for Fig. \ref{compro}, the scaled $R<2$ kpc mass at $z=0$ is for all samples lower than in observed galaxies.
D13 found $\sim 1.5 \times 10^{10}\;M_{\odot}$, whereas we find a scaled $R<2$ kpc mass of $\sim 0.9\times 10^{10}\;M_{\odot}$ for all galaxies
and $\sim 0.6\times 10^{10}\;M_{\odot}$ for the MW mass sample.
We note that, after scaling, the factors of growth from $z=2.5$ to $z=0$ for all samples change in a way that yields better agreement with
observed growth factors. Moreover, the evolution curves now show flattening at late times, as those found by D13, although central 
growth flattens too early. This exercise shows that the discrepancies are in part caused by the difference between the SFH of our
model galaxies and that inferred for the observed sample, as visualized in Fig. \ref{massevo}.

\begin{figure}
\centering
\includegraphics[width=9cm]{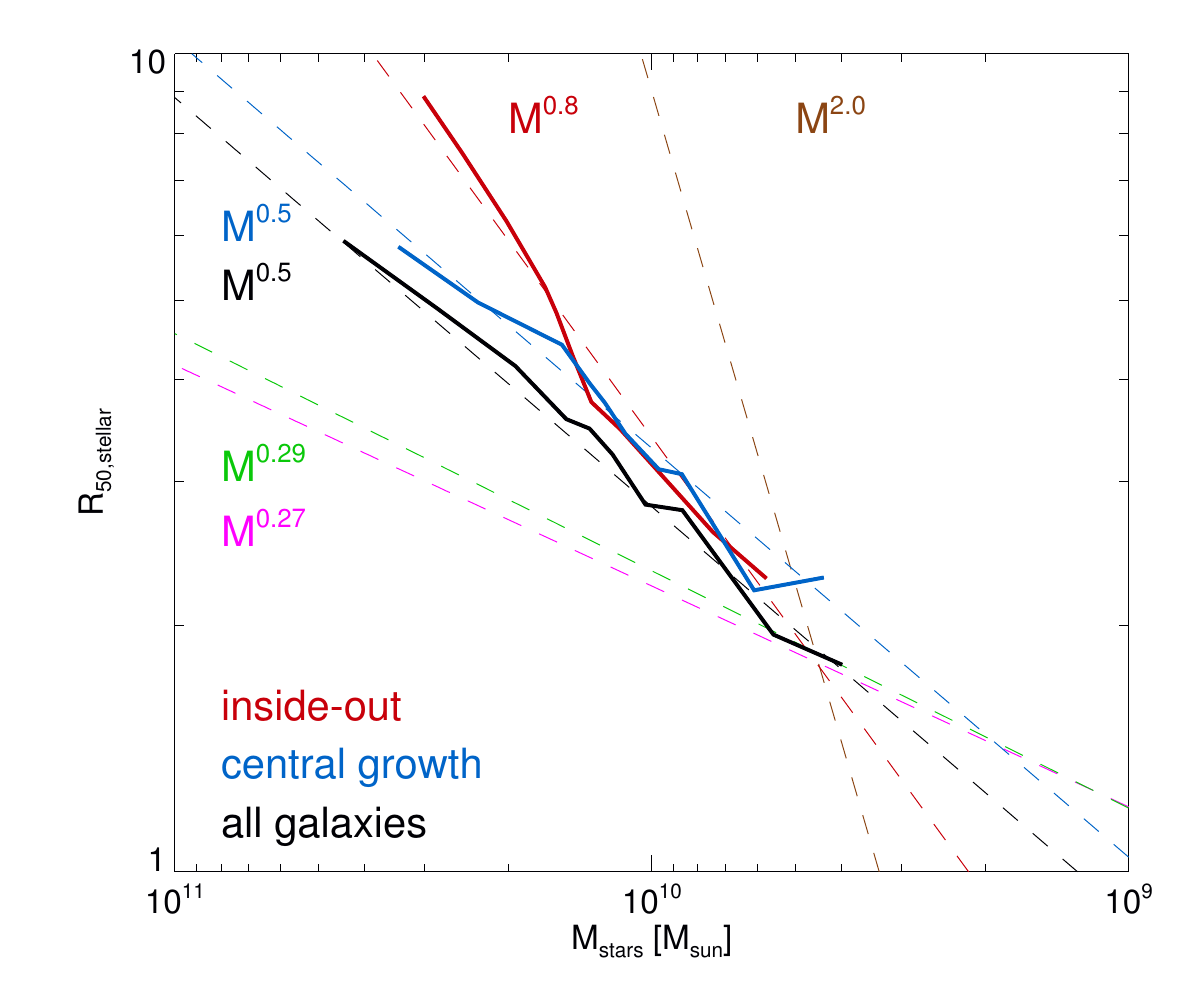}
\caption[The evolution of simulated galaxies in the mass-size plane]
{The evolution of simulated galaxies in the plane of stellar half-mass radius $R_{50, \rm stellar}$ and stellar mass $M_{\rm stellar}$.
 The all galaxies (black) and central growth (cyan) samples follow roughly an $R\propto M^{0.5}$ growth, whereas the inside-out sample
 (red) follows approx. an $R\propto M^{0.8}$ growth. For comparison the results for disc galaxies by D13 (magenta $R\propto M^{0.27}$)
 and P13 (green $R\propto M^{0.29}$) and for elliptical galaxies by \citet{patel2} (brown $R\propto M^{2.0}$) are shown.
 The latter was shifted to lower masses to allow a comparison between the relations.
 The mean mass evolution of the simulated samples is used for scaling.}
\label{evols2}
\end{figure}

Another structural property that encodes the shape of the profile is the Sersic index $n_S$. P13 and D13 displayed that the galaxies they observed
showed an increase from $n_S\sim 1$ at high $z$ to $n_S\sim 2.3-3$ at $z=0$ for one-component fits. We also perform one-component Sersic
fits to the mean profiles of the four samples introduced above and depict the results in the right panels of Fig.
\ref{evols}. Fits can depend significantly on the binning of the data or on the fitted region and the uncertainty is thus high.
Independent of the assumed scaling and of sample selection,
we find a mild increase of Sersic index, from $n_S\sim 1.3-1.6$ at $z>2$ to $n_S\sim 1.8-2.4$ at $z=0$. If we redo the 
procedure with $g$ band surface brightness profiles instead of mass profiles, we find somewhat lower indices at high $z$ and higher
indices at low $z$. This is especially true for the inside-out sample, for which the low $z$ profiles show clear disc and bulge regions
(see AqC in Fig. \ref{radial-g} for an extreme example) and are generally not well fit by one-component profiles.
Considering these trends, the agreement with observations is reasonable.

The biggest discrepancies between observations and simulations become clear in the middle panels of Fig. \ref{evols}, where we plot
the evolution of stellar half-mass radii. In A13 we already showed that the simulated galaxies tend to be too extended at $z=0$
compared to the mass-size relation found by \citet{shen03} for local galaxies. This discrepancy was biggest for
the galaxies with MW-like masses. From above, we have learned that this is driven by overly low central masses.
D13 showed that their galaxies had half-light radii of $R_{50}\sim 2$ kpc at $z>2$, which only increased mildly to $R_{50}\sim 3.5$ kpc
at $z=0$ with hardly any evolution after $z=1$. The results of P13 are very similar.

Although for all our different samples of simulated disc galaxies we also find sizes $R_{50}\sim 2$ kpc at $z>2$, the increase
in size is much stronger at later times, especially at $z<1$.
Assuming the mean mass evolution of each sample for the scaling procedure,
the final mean half-mass radius is $R_{50}\sim 5.8$ kpc for all galaxies, $R_{50}\sim 5.9$ kpc for the 
central growth sample, $R_{50}\sim 7.1$ kpc for the MW mass sample and $R_{50}\sim 8.8$ kpc for the inside-out sample.
Size thus clearly anti-correlates with central mass growth. We also overplot half-light radii for our mock $g$ band profiles
which are again $\sim 1/3$ more extended at $z=0$ than the corresponding half-mass radii.
When the mass evolution selected by D13 is assumed for the scaling, radii also increase by $\sim 10\%$.

In Fig. \ref{evols2} we plot the evolution of the samples in the mass-size plane. P13 and D13 find for their observational samples
relations similar to $R_{50}\propto M^{0.3}$, whereas \citet{patel2} found $R_{50}\propto M^{2.0}$ for elliptical galaxies.
At high $z$, all our samples start to evolve as predicted for observed disc galaxies, but then evolve away from this relation.
We find that our all galaxies and central mass growth samples grow as $R_{50}\propto M^{0.5}$ and our inside-out sample
grows as $R_{50}\propto M^{0.8}$. The growth is thus significantly stronger than for observed disc galaxies, but significantly
weaker than for observed ellipticals.  Scaling according to the mean mass evolution of the simulated samples was used
for Fig. \ref{evols2}, but the slopes are independent of scaling. \citet{hirschmann} recently presented this plot for a larger
set of lower resolution resimulations with a very different feedback model compared to ours. Interestingly, they find 
a better agreement for the size evolution ($R_{50}\propto M^{0.4}$ for their WM feedback model and MW mass galaxies),
although their model galaxies are at all times too compact.

In summary, we have shown that the structure of our simulated galaxies agrees with that of the observed galaxies at $z\ge1.5$, independent 
of sample construction. After $z=1$, all galaxies, especially those with $z=0$ stellar masses around $1-6\times 10^{10}M_{\odot}$,
grow a factor of $\sim 2$ too little in central mass and consequently a factor of $\sim 2$ too much in size. The discrepancy with
observations is strongest for inside-out growing galaxies, but also valid for other formation histories.
Discrepancies at low $z$ are stronger for mock light profiles than for mass profiles.
We also find that this comparison between observations and simulations of disc galaxy evolution is complicated
by the variety of SFHs of simulated discs, which often differ from simple assumptions chosen for observations.

\section{Conclusions and Discussion}

We have presented an analysis of the evolution of the structural properties of a sample of 19 simulated disc galaxies.
These are cosmological re-simulations with multiphase SPH which were recently presented by A13. 
We have created mock three-colour $u/g/r$ band images to visualize the evolution
of the models. We have investigated, how the complexity of light profiles and variations with orientation 
can complicate the interpretation of observed data if rest-frame $g$ or $i$ band light is used as a proxy for mass, as in the work of D13 or P13.
The model galaxies show a variety of structural evolution histories, ranging from inside-out growing discs around
older, more centrally concentrated stellar populations to continuous SF at all radii.

Our models show only a small number of bars and no unstable, clumpy discs, so that not all known mechanisms for central mass growth
in disc galaxies are well represented. However, gas-rich mergers with mass fractions up to $1:3$ 
and misaligned infall of gas can be clearly linked to episodes
of central mass growth. In a $\Lambda$CDM universe, these events are common and are thus likely to be major contributors
to the continuous central growth observed in mean surface density profiles of samples of galaxies selected at various
redshifts in order to represent the typical evolution history of galaxies with $z=0$ masses similar to that of the MW (D13 and P13).
Further observational evidence for such a scenario was recently presented by \citet{kaviraj}, who estimated that SFR enhancements in morphologically
disturbed disc galaxies contribute significantly to present-day SF in spiral galaxies. These disturbed galaxies also exhibit
central mass growth and are plausibly undergoing minor mergers or misaligned infall as discussed above.

We have also shown that model galaxies which have undergone mergers or misaligned infall events after $z=1$ can have
very low values of $B/T<0.1$ for the bulge-to-total ratio if mock face-on $i$-band profiles are decomposed into an exponential
disc plus a Sersic bulge profile. Despite their appearance as good disc galaxies, their kinematic disc fractions are often
below 50 per cent (see also \citealp{cs10}). Observed galaxies with low photometric $B/T$ might thus not all 
have had quiescent merger histories.

Model galaxies which do not experience mergers  or misaligned infall events, grow inside-out and show no or little central mass growth.
We have shown that mergers or misaligned infall leave distinct features in the distributions of present-day radii, circularities and metallicities
as functions of stellar age. Observational detection of such features is complicated by the fact that the timescales of
these events are smaller than the typical errors in age determinations of stars.

In our simulations, metallicities effectively increase during events that drive central SF, as these events enrich the ISM. 
The evolution of metallicities in discs due to interactions has been studied in detail by several groups \citep{montuori, perez, torrey}.
All these papers discuss non-cosmological simulations of mergers of pairs of disc galaxies. They all apply SPH codes
without a prescription for the turbulent diffusion of metals, which is included in our fully cosmological models and has been shown to 
significantly change the metallicity distributions in discs \citep{pilkington} and the circumgalactic medium \citep{shen}.
All these models find a strong initial dilution of central metallicities caused by the inflow of gas originating from the outer disc
which, due to the metallicity gradients assumed in the set-up, is metal poor. In A13 we showed 
that the discs in our simulations have gradients that are shallower
than observed and \citet{Gibson} recently showed that the details of feedback models significantly influence gradients.
Because of the shallow gradients, only very weak metallicity dilution features in the form of slightly lowered metallicities 
at the beginning of the events can be found in Fig. \ref{archae}.
For the idealized simulations, it has, however, also been shown that enrichment due to SF in gas-rich mergers leads to an effective
increase of metallicities at the end of the merger, in agreement with our findings. We conclude that events which trigger
central SF are likely to leave an imprint on the age-metallicity relation of stars in that region
and possibly also throughout the whole disc, but the shape of this feature is uncertain on account of modelling uncertainties.

When we compare our sample of simulated galaxies directly to observed samples, we have to take into account that
the allowed scatter in the observers' selection criteria on evolution histories is much narrower than the scatter of the evolution histories
displayed by our simulated galaxies. Moreover, the shapes of the SFHs of our galaxies differ significantly from that assumed by D13,
although in A13 we have shown that the models with similar $z=0$ masses as targeted by D13 agree well
with abundance matching predictions for SFHs by \citet{moster}. 
Agreement with the assumptions of P13 in terms of SFH at slightly lower galaxy masses is better.

Nevertheless, independent of sample selection, the mean profiles of the simulated galaxies agree with observations
at $z\ge1.5$. At later times, galaxies show too little central mass growth and effective radii that increase too rapidly.
At $z=0$ galaxies are typically a factor of $\sim2$ too extended. Disagreement at low $z$ is significant both for mass
and light profiles and strongest if $g$-band light is used.
It is not, however, clear how much this result is influenced
by the biased selection of haloes, as a significant number of galaxies with effective radii $R_{50}\sim 10$ kpc have been observed
\citep{sargent}. Moreover, in a comparison of our $z=0$ model gas discs with observed HI discs, we find that the observed
and simulated samples show the same range of gas disc diameters \citep{wang} and follow the same HI mass-size relation.

Size growth is strongest for galaxies with masses similar to that of the MW and several of those models show strong inside-out growth.
Although our full sample of galaxies shows better agreement with observations, we caution that this behaviour is driven
by the high central SFRs and high central stellar mass densities at $z=0$ of the most massive galaxies,
which are likely unrealistic. \citet{fang} recently showed that low-redshift galaxies with high central mass densities
are preferentially quenched. The lack of a model of AGN feedback in our simulations is a possible cause for this.

It is interesting to note that our models agree best with observations at high $z$, as, until recently, simulations matched observational data
particularly poorly at early formation stages (e.g. \citealp{aquila,moster}).
As already discussed in A13, our new simulations show better agreement
with a variety of observations at $z\sim 1$ than at $z=0$. One clearly identified problem is the lack of bars, which predominantly form
at low $z$ and can lead to inflow of gas to the centres of discs.
One reason for this are overly high velocity dispersions in our model discs (see A13), which can prevent the 
formation of dynamically cold structures. Coarse resolution
prevents low velocity dispersions (see e.g. \citealp{house}), but other simulations
at the same resolution have produced many barred systems (e.g. \citealp{cs09}).
Our models differ from these as they have lower surface densities and lower baryon conversion efficiencies.

The main reason for the differences is the strong stellar feedback applied in our models. 
As was shown in \citet{hannah}, our feedback implementation is very efficient at removing low angular momentum material.
In general, the angular momentum of accreted gas in our models also increases with time.
Moreover, low angular momentum gas that is blown out in early formation stages can re-accrete after several Gyr,
but then has on average gained angular momentum. All these points combine to keep central baryonic surface densities
low and to reduce bulge fractions. They thus also help in preventing disc instabilities, which could lead
to bar formation and thus enhance central surface densities, which in turn would make the discs more unstable due to the
increase in surface gravity.

\citet{marinacci} recently simulated five of our haloes with a different galaxy formation code. Interestingly,
in these haloes, they found more massive and 
less extended galaxies compared to our models. On average, their sizes lie mildly above the observed mass-size relation of \citet{shen03}.
They used a very different prescription for stellar feedback and also find a higher fraction of bars.

These arguments seem to suggest that the effects of the stellar feedback models applied in our simulations are too strong at low redshifts,
as was also indicated by the underproduction of stars at low $z$ in lower mass galaxies, as discussed in A13.
It is interesting that the model for feedback from radiation pressure from young massive stars already invokes a significantly stronger
feedback for high $z$ gas-rich, turbulent galaxies than for quiescently star-forming discs at low $z$.
It is possible that the calibration on galaxies similar in mass to the MW is problematic. At this halo mass the quenching 
mechanism needed for high mass galaxies, be it AGN or something else, but not implemented here, 
probably already plays a role. Moreover, in our simulations, feedback is particularly efficient in removing material during
mergers \citep{hannah}, so that central mass growth during these events might actually be underestimated in our models.

Simulations of disc galaxy formation have long been struggling with removing sufficient amounts of low angular momentum 
material from the inner regions and avoiding the formation of overly compact galaxies. Apparently our efforts in A13 to overcome this
have led to the opposite problem. Weakening feedback seems an obvious solution, but the success of the model
at high redshifts suggests that this should only be effective at late times.
Evidently, our models are still far from capturing all relevant astrophysical processes accurately.

\section*{Acknowledgments}

We thank the referee, Vladimir Avila-Reese, for helpful comments.
MA and TN acknowledge support from the DFG Excellence Cluster "Origin and Structure of the Universe".
SDMW was supported in part by Advanced Grant 246797 "GALFORMOD" from the European Research Council.

\end{document}